\documentclass[a4paper,prb,reprint,showkeys,floatfix]{revtex4-1}

\usepackage[latin1]{inputenc}
\usepackage[english]{babel}
\usepackage{amsmath}
\usepackage{ae}
\usepackage{color}
\usepackage{graphicx}
\usepackage{subfigure}
\usepackage{hyperref}
\hypersetup{
  colorlinks   = true, 
  urlcolor     = blue, 
  linkcolor    = blue, 
  citecolor   = blue 
}

\def\vk{\mbox{$\vec{k}$}}

\newcommand{\pmat}[4]{\ensuremath{\begin{pmatrix} #1 & #2\\ #3 & #4 \end{pmatrix}}}
\newcommand{\pvec}[2]{\ensuremath{\begin{pmatrix} #1\\ #2 \end{pmatrix}}}

\begin{document}

\title{Optical signatures of nonlocal plasmons in graphene}

\author{Tobias Wenger}
\author{Giovanni Viola}
\author{Mikael Fogelstr\"om}
\affiliation{Department of Microtechnology and Nanoscience (MC2), Chalmers University of Technology, S-412 96 G\"oteborg, Sweden}
\author{Philippe Tassin}
\author{Jari Kinaret}
\affiliation{Department of Physics, Chalmers University of Technology, S-412 96 G\"oteborg, Sweden}

\date{\today}

\begin{abstract}
We theoretically investigate under which conditions nonlocal plasmon response in monolayer graphene can be detected. To this purpose, we study optical scattering off graphene plasmon resonances coupled using a subwavelength dielectric grating. We compute the graphene conductivity using the Random Phase Approximation (RPA) obtaining a nonlocal conductivity and we calculate the optical scattering of the graphene-grating structure. We then compare this with the scattering amplitudes obtained if graphene is modeled by the local RPA conductivity commonly used in the literature. We find that the graphene plasmon wavelength calculated from the local model may deviate up to $20\%$ from the more accurate nonlocal model in the small-wavelength (large-$q$) regime. We also find substantial differences in the scattering amplitudes obtained from the two models. However, these differences in response are pronounced only for small grating periods and low temperatures compared to the Fermi temperature.

\end{abstract}


\maketitle


\section{Introduction}
Monolayer graphene has attracted much attention due to its remarkable electronic and optical properties \cite{Peres2006,Stauber2008,CastroNeto2009,Bonaccorso2010,DasSarma2011}. For instance, monolayer graphene has broadband absorption of $2.3\%$ \cite{Nair2008}, which is quite substantial since graphene is only one atom thick. The doping level in monolayer graphene is also tunable by applying external gating \cite{Novoselov2004} and it exhibits an optical response ranging from terahertz to optical frequencies \cite{Grigorenko2012}. The exciting properties of graphene arise from a combination of its 2-dimensional nature and its hexagonal lattice structure. Together these properties make the low-energy electrons obey an effective massless Dirac equation \cite{katsnelson}, which also has consequences for the collective plasmon excitations in graphene.\\

Plasmons in graphene have been known for quite some time \cite{Shung1986,Gusynin2006,Wunsch2006,Hwang2007} and exhibit strong confinement of the electromagnetic fields \cite{Jablan2009}. The plasmon wavelength is much smaller than the free-space wavelength of light, for instance making it possible to achieve subwavelength resolution microscopy \cite{Gramotnev2010}, and facilitates strong light-matter interaction \cite{Koppens2011}. Furthermore, the strong field confinement of graphene plasmons has recently been used for ultra-sensitive detection of molecules \cite{Rodrigo2015,Farmer2016}. Other proposed applications of graphene plasmons include modulators, filters, polarizers, and photodetectors \cite{Low2014,Bao2012}.\\

However, due to their small wavelength it is challenging to interact with and to detect graphene plasmons and many different schemes have been proposed. Examples include introducing metal antennas on top of the graphene surface \cite{Gonzalez2014}, patterning the graphene into microribbon arrays\cite{Ju2011,Wang2012a,Yan2013} or microdisk arrays \cite{Thongrattanasiri2012a,Yan2012}, using total internal reflection \cite{Ukhtary2015}, introducing a periodic spatial modulation of the graphene conductivity \cite{Peres2012,Slipchenko2013}, and placing a nano-tip close to the graphene surface \cite{Chen2012,Fei2012,Woessner2015b,Torre2015,Gonzalez2016,Lundeberg2016}. Another approach is to pattern the substrate into a grating \cite{maier} which has been experimentally demonstrated in Refs. \cite{Zhu2013,Jadidi2015}.\\

\begin{figure}[h!]
\centering
\includegraphics[width=0.4\textwidth]{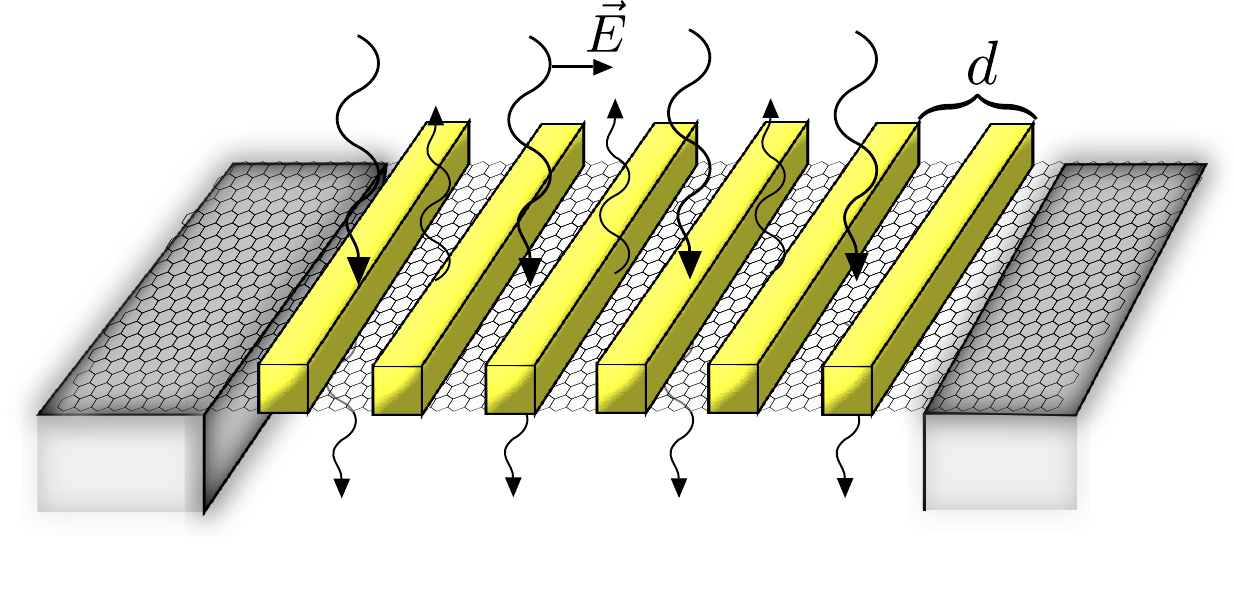}
\caption{\label{fig:setup}(Color online) Grating on top of graphene and incident radiation with the electric field in the longitudinal direction, i.e., perpendicular to the grating. The incident electric field has amplitude $E_0$, the reflected field amplitude is $rE_0$ and the transmitted field amplitude is $tE_0$.}
\end{figure}

In this paper, we theoretically investigate the optical scattering of a system consisting of a subwavelength dielectric grating and a doped monolayer graphene sheet, as shown in Fig.~\ref{fig:setup}. The scattering amplitudes are computed using a scattering matrix method \cite{Li2003} and the graphene enters our electromagnetic problem as a conducting boundary condition. We calculate the graphene conductivity using the Random Phase Approximation (RPA), yielding a nonlocal conductivity $\sigma(q,\omega)$. A local expression can be obtained by taking the limit $\sigma(q\to0,\omega)$---this is usually called the local RPA result in literature. The combined system of graphene together with a subwavelength dielectric grating has been treated previously \cite{Zhan2012,Gao2012,Liu2013,Peres2013} using the local RPA. The local RPA result is expected to correctly describe long-wavelength plasmons (where $q$ is small), but since much interest in plasmons arise from their small wavelengths (where  $q$ is large), it is important to also investigate nonlocal effects.\\

The conductivity of graphene has been the subject of much research lately and in particular the effects of disorder \cite{Kechedzhi2013,Principi2013}, phonons \cite{Jablan2009,Principi2014} and electron-electron interaction \cite{Polini2008,Kechedzhi2013} have been investigated using various approaches. Plasmon-phonon hybridization has also been experimentally investigated in Refs.~\cite{Koch2010,Koch2016}. We assume that we are far from resonance with any phonon in our system and we also assume that our samples are clean enough to neglect impurities. We also neglect electron-electron interaction effects in our treatment. Our focus will be on quantifying the nonlocal effects (non-zero $q$) by comparing with the local RPA. Previous studies of nonlocal effects in graphene include Refs.~\cite{Wang2013,Thongrattanasiri2012b}. It was found that nonlocal effects influence the plasmon dispersion and also the plasmon width in both nanoribbons and nanodisks. We also investigate the temperature dependence of our results.\\

The paper is organized as follows; in section \ref{sec:plasmonMode} we calculate the plasmon dispersion and quantify the intrinsic plasmon width. Section \ref{sec:resonancePeaks} contains our calculated results for the reflectance, transmittance and absorbance in the combined system of graphene and subwavelength dielectric grating for one specific grating. In section \ref{sec:gratings} we investigate the optical response for various grating periodicities and temperatures.

\section{Longitudinal surface plasmon modes}\label{sec:plasmonMode}

In order to study the plasmons, it is convenient to calculate the plasmon dispersion, i.e., the relationship between energy and momentum of the plasmon mode. This has previously been studied at zero temperature in Refs.~ \cite{Hwang2007,Wunsch2006} and at finite temperature in Refs. \cite{DasSarma2013,Jablan2009,Wang2012b}. Having a conductor in between two dielectrics, an equation for modes confined to the conductor can be derived from Maxwell's equations \cite{Jablan2009}:

\begin{equation}\label{eq:MEdisp}
\frac{\epsilon_{\uparrow}}{\sqrt{q^2-\frac{\omega^2\epsilon_{\uparrow}}{c^2}}}+\frac{\epsilon_{\downarrow}}{\sqrt{q^2-\frac{\omega^2\epsilon_{\downarrow}}{c^2}}}+\frac{i\sigma(q,\omega)}{\omega\epsilon_0}=0
\end{equation}
where $q$ is the in-plane wave vector, $\sigma(q,\omega)$ is the sheet conductivity of graphene, and $\epsilon_{\uparrow}$ and $\epsilon_{\downarrow}$ are the relative dielectric constants above and below the graphene sheet. Solving the real part of Eq. \eqref{eq:MEdisp} for $\omega$ as a function of $q$, we obtain the plasmon dispersion. The nonlocal conductivity of graphene is computed using linear response theory (details are given in appendix \ref{app:linResp}).\\

Another convenient way to investigate intrinsic plasmon properties is to calculate the spectral function of density fluctuations \cite{mahan,vignale}
\begin{align}\label{eq:lossFunc}
S(q,\omega)&=-\frac{1}{v_q}\text{Im}\left [\frac{1}{\epsilon(q,\omega)} \right ]=\nonumber\\
&=\frac{1}{v_q}\frac{\epsilon_2(q,\omega)}{\epsilon_1(q,\omega)^2+\epsilon_2(q,\omega)^2}
\end{align}
where $\epsilon(q,\omega)=\epsilon_1(q,\omega)+i\epsilon_2(q,\omega)$. The RPA expression for the dielectric function is \cite{mahan,vignale}
\begin{equation}
\epsilon(q,\omega)=1-v_q\Pi(q,\omega)
\end{equation}
where $v_q=\frac{e^2}{q\epsilon_0(\epsilon_{\uparrow}+\epsilon_{\downarrow})}$ and $\Pi(q,\omega)$ is the polarizability. For a definition of the polarizability see appendix \ref{app:linResp}.\\

In order to relate equations \eqref{eq:MEdisp} and \eqref{eq:lossFunc}, we rewrite Eq.~\ref{eq:MEdisp} using the so called non-retarded approximation, i.e., $q\gg\omega/c$. Since we are interested in the plasmon behavior of strongly localized plasmons this is a valid approximation. Eq. \ref{eq:MEdisp} then becomes
\begin{equation}\label{eq:temp1}
1+\frac{iq}{\epsilon_0(\epsilon_{\uparrow}+\epsilon_{\downarrow})\omega}\sigma(q,\omega)=0
\end{equation}
and from appendix \ref{app:linResp} we have
\begin{equation}
\sigma(q,\omega)=\frac{ie^2\omega}{q^2}\Pi(q,\omega).
\end{equation}
Inserting this in Eq.~\eqref{eq:temp1}, we get 
\begin{equation}
1-v_q\Pi(q,\omega)=0
\end{equation}
which by definition is equivalent to
\begin{equation}
\epsilon(q,\omega)=0.
\end{equation}
This equation is often used to determine the plasmon dispersion in the literature. Here we see it emerging as a non-retarded approximation to Eq. \eqref{eq:MEdisp}.\\

The width of the surface plasmons can be estimated by substituting the definition of the dielectric function into Eq.~\eqref{eq:lossFunc} for the spectral function, yielding
\begin{equation}
S(q,\omega)=\frac{-\Pi_2(q,\omega)}{(1-v_q\Pi_1(q,\omega))^2+v_q^2\Pi^2_2(q,\omega)}
\end{equation}
where $\Pi_1$ ($\Pi_2$) is the real (imaginary) part of $\Pi$. Close to the plasmon frequency $\omega_p$, $\epsilon_1(q,\omega)=1-v_q\Pi_1(q,\omega)$ may be expanded as $\epsilon_1(q,\omega)\approx -v_q(\omega-\omega_{\text{p}})\partial_{\omega}\Pi_1(q,\omega)|_{\omega=\omega_{\text{p}}}$ and inserting this into the spectral function we obtain
\begin{align}
S(q,\omega)=I_0(q,\omega)\frac{\gamma(q,\omega)^2}{(\omega-\omega_{\text{p}})^2+\gamma(q,\omega)^2}
\end{align}
where we have defined
\begin{align}\label{eq:gammaLoss}
\gamma(q,\omega)&=\frac{\Pi_2(q,\omega)}{\partial_{\omega}\Pi_1(q,\omega)|_{\omega=\omega_{\text{p}}}}\\
I_0(q,\omega)&=-\frac{1}{v_q\Pi_2(q,\omega)}.
\end{align}
Close to the plasmon frequency, this resembles a Lorentzian with height $I_0(q,\omega)$ and half width at half maximum (HWHM) $\gamma(q,\omega)$. This is strictly only true if $I_0(q,\omega)$ and $\gamma(q,\omega)$ are constant close to the plasmon resonance.\\

\begin{figure}[h!]
\centering
\subfigure[]{\label{fig:dispTemp}
\includegraphics[width=0.4\textwidth]{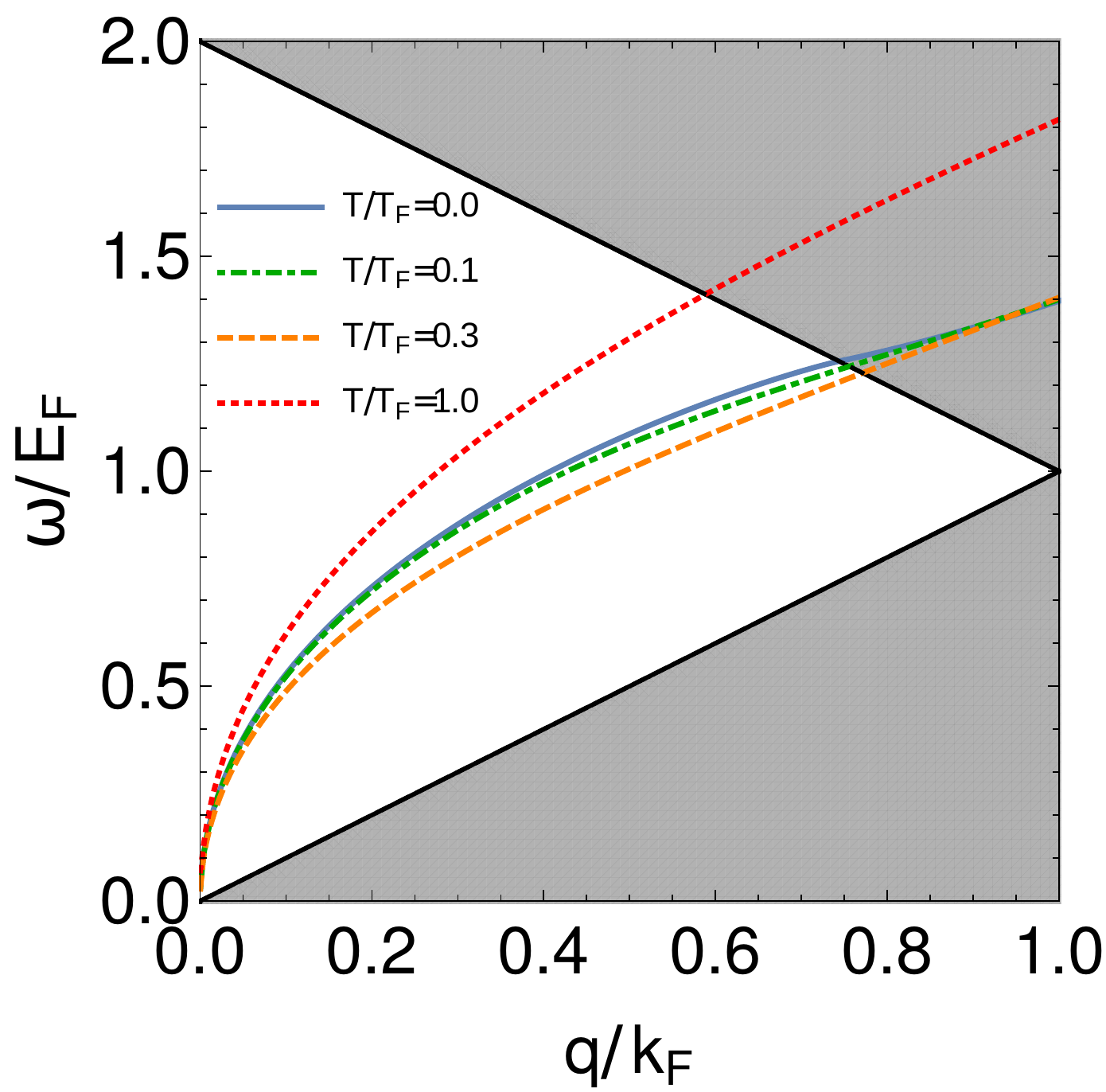}
}
\subfigure[]{\label{fig:dispLossK}
\includegraphics[width=0.4\textwidth]{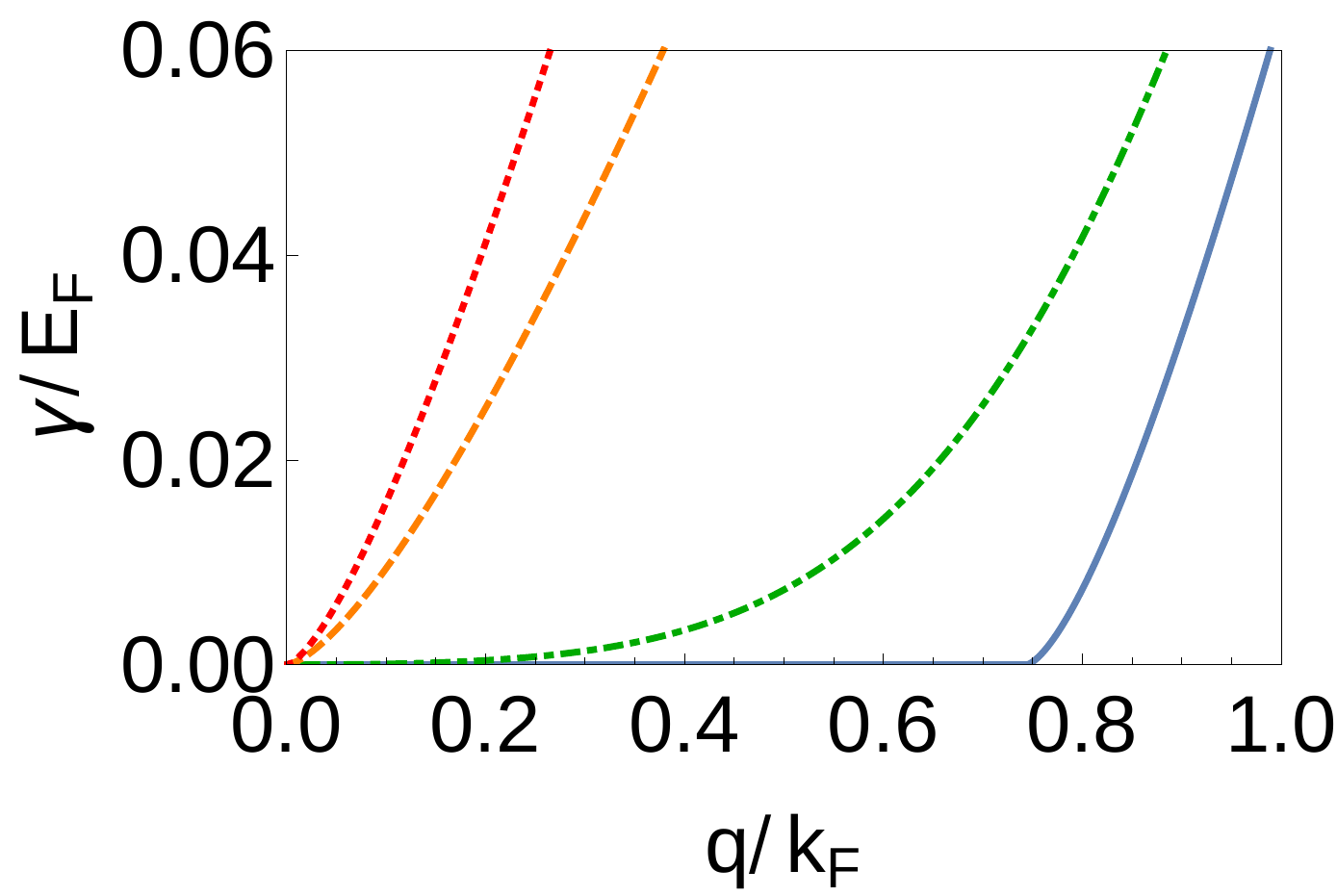}
}
\caption{\label{fig:plasmonMode}(Color online) \textbf{a)} Plasmon dispersions for different temperatures obtained by solving the real part of Eq.~\eqref{eq:MEdisp}. The graphene has vacuum on one side and a dielectric substrate with $\epsilon_r=2$ on the other side. The white triangle is defined by the real part of $\sigma(q,\omega)$ being identically zero at zero temperature due to Pauli blocking of interband transitions. \textbf{b)}Intrinsic plasmon width of the dispersions in a) obtained from Eq.~\eqref{eq:gammaLoss}. In both figures $T/T_F=0.0$ (blue solid line), $T/T_F=0.1$ (green dot-dashed line), $T/T_F=0.3$ (orange dashed line) and $T/T_F=1.0$ (red dotted line).}
\end{figure}

Fig. \ref{fig:dispTemp} shows the plasmon dispersion for four different temperatures obtained by solving the real part of Eq.~\eqref{eq:MEdisp} using $\sigma(q,\omega)$\footnote{As we have shown the dielectric function is equivalent to Eq.~\eqref{eq:MEdisp} in the non-retarded approximation. This means that in order to obtain the plasmon dispersion we could just as well solve for zeros to $\epsilon_1(q,\omega)$ and the only difference would be that the dispersions would be slightly altered for small $q/k_F$. However, these differences are too small to observe in Fig.~\ref{fig:dispTemp}. This could however be important when investigating plasmons with larger wavelength (small $q/k_F$).}. Temperature effects on the plasmon dispersion are modest at small $T/T_F$, but shifts the dispersion curve significantly at temperatures $T/T_F\approx1$. We clearly observe a non-monotonic behavior for the dispersion as a function of the temperature, which was previously discussed in Ref. \cite{Ramezanali2009}. In Fig.~\ref{fig:dispTemp}, we wish to clarify the white triangle. This area is defined by the real part of $\sigma(q,\omega)$ [or imaginary part of $\Pi(q,\omega)$] being identically zero at zero temperature due to Pauli blocking of interband transitions. The spectral function then becomes a delta function, as the width of the Lorentzian goes to zero, and the plasmon mode is an infinitely sharp energy state. However, this is only true in the absence of impurities and at zero temperature. Adding impurities and/or going to non-zero temperature make $\Pi_2(q,\omega)$ non-zero, resulting in a non-zero plasmon width. Below we investigate how the plasmon width is affected by non-zero temperatures.\\

Fig.~\ref{fig:dispLossK} shows the intrinsic plasmon width obtained from Eq.~\eqref{eq:gammaLoss} for the dispersions in Fig.~\ref{fig:dispTemp}. We see that the zero-temperature plasmon width is indeed zero inside the white triangle in Fig.~\ref{fig:dispTemp}. The zero-temperature plasmon obtains a non-zero width for $q/k_F\gtrsim0.74$, which is where the zero-temperature dispersion crosses from the white triangle to the gray shaded area in Fig.~\ref{fig:dispTemp}. The temperature can be seen to affect the high energy plasmons (large $q/k_F$) more than the low energy plasmons (small $q/k_F$). However, when the temperature is of the order of the Fermi temperature, the low energy plasmons are affected too.\\

\begin{figure*}
\centering
\subfigure[]{\label{fig:dispCompare}
\includegraphics[width=0.3\textwidth]{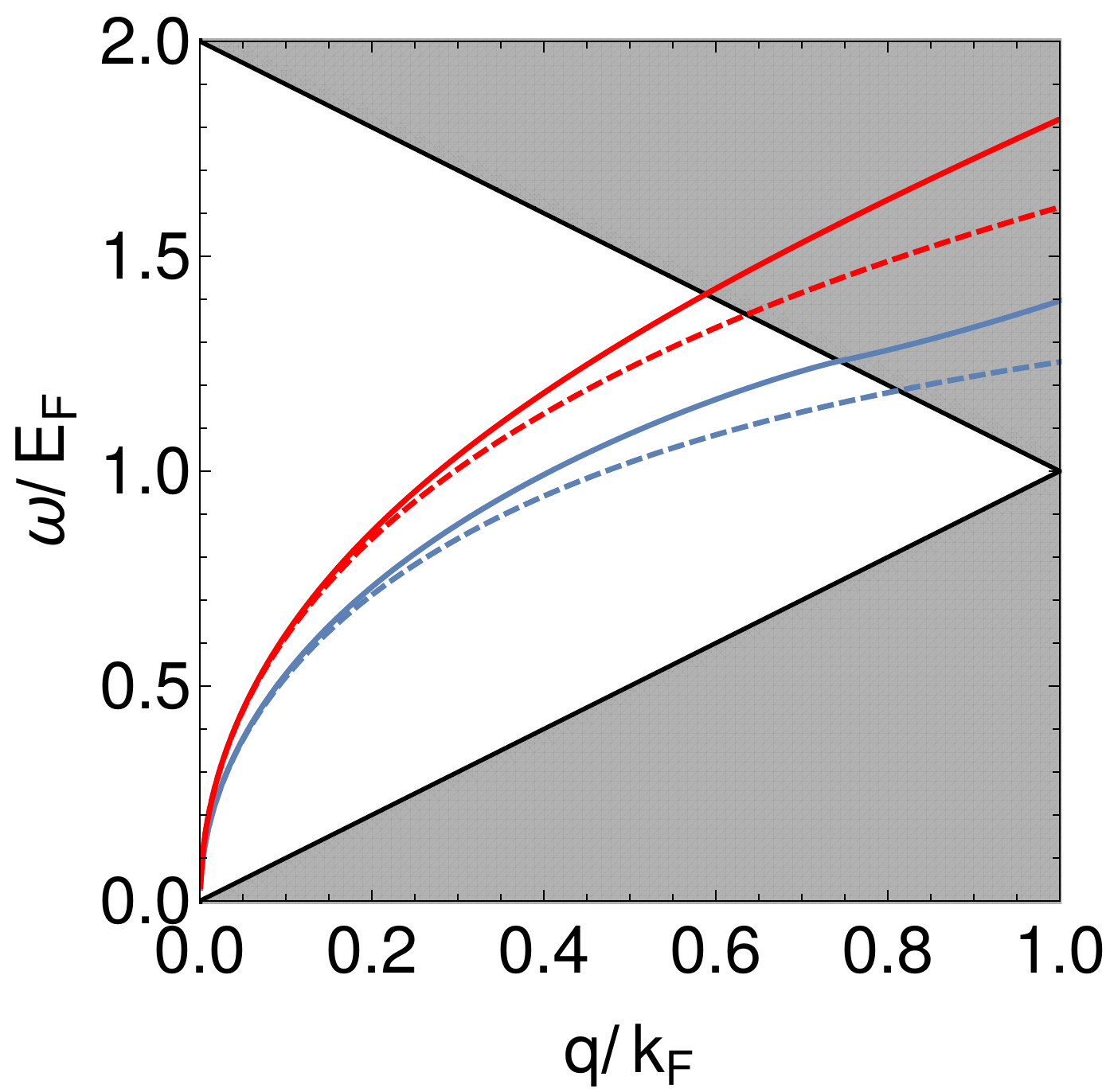}
}
\subfigure[]{\label{fig:wavelength}
\includegraphics[width=0.3\textwidth]{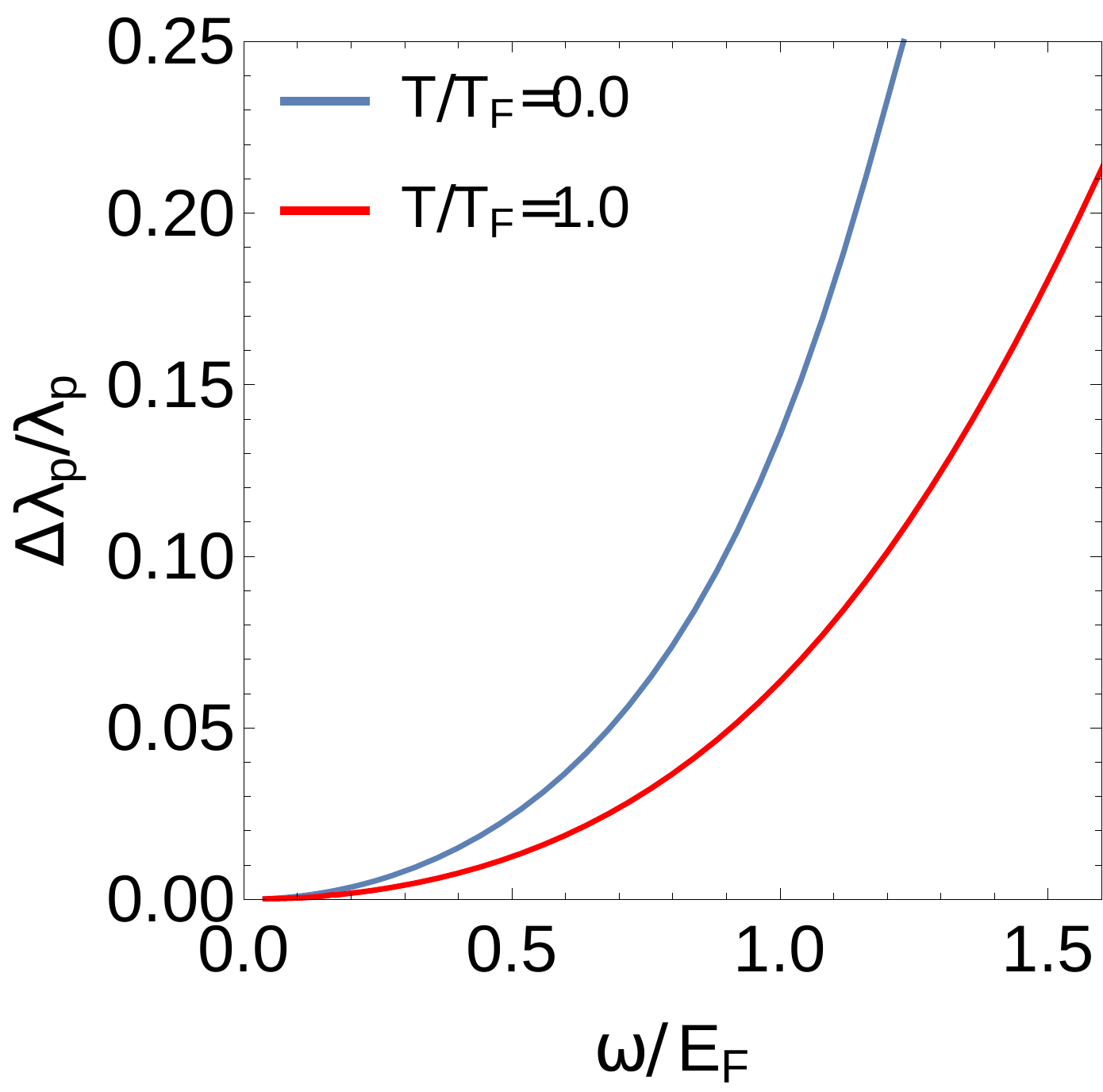}
}
\subfigure[]{\label{fig:diffLoss}
\includegraphics[width=0.3\textwidth]{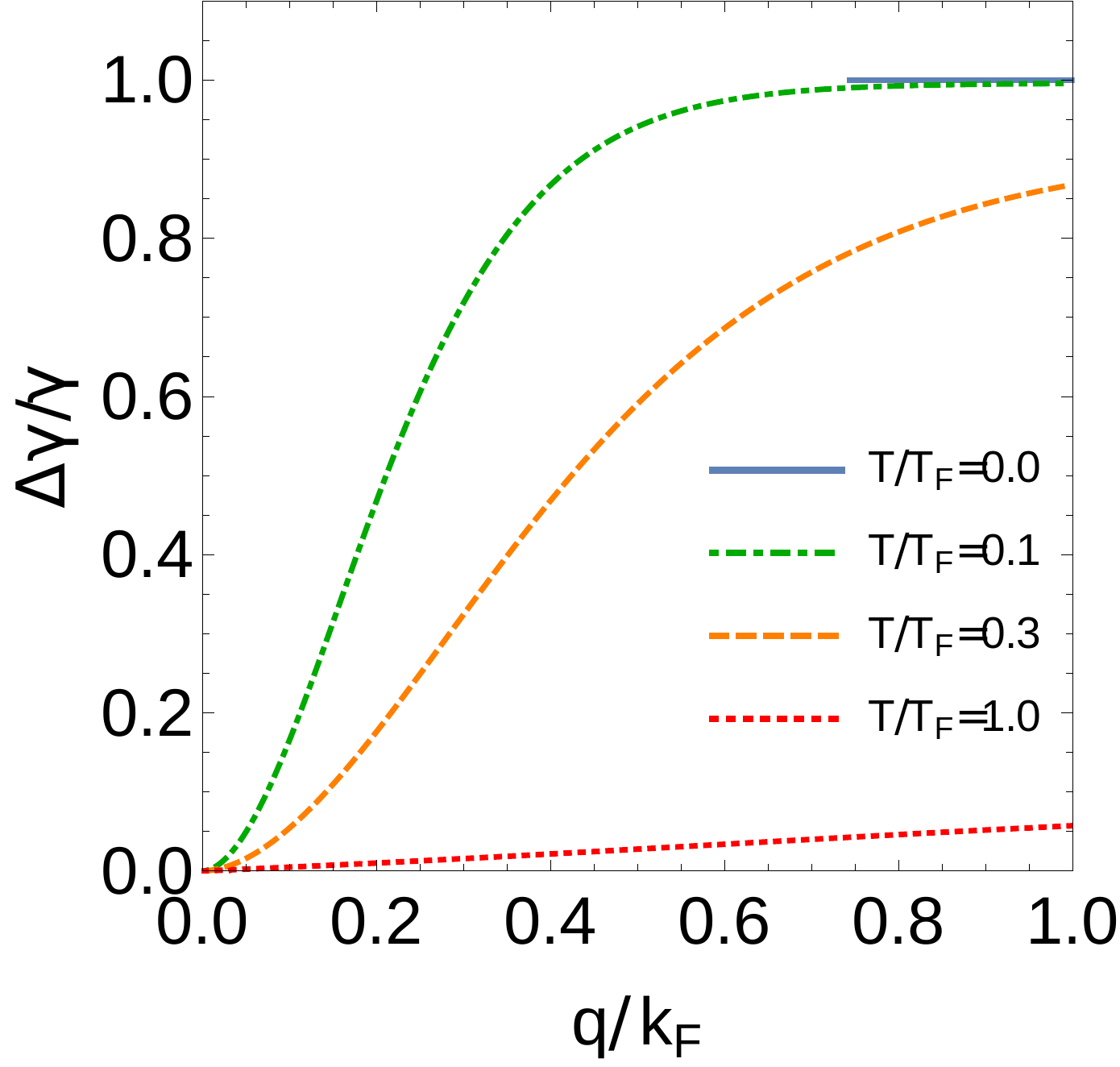}
}
\caption{(Color online) \textbf{a)} Dispersion relations illustrating the differences between using nonlocal RPA (solid lines) and local RPA (dashed lines). The blue lines are obtained for $T/T_F=0.0$ and the red lines for $T/T_F=1.0$. \textbf{b)} The differences between the surface plasmon wavelengths obtained from the dispersions in a). The blue line is the relative difference between the results at $T/T_F=0.0$ and the red line is the relative differences between the result at $T/T_F=1.0$. \textbf{c)} Relative difference in the plasmon width obtained from Eq.~\eqref{eq:gammaLoss} using nonlocal RPA and local RPA. The colors represent the same colors as in Fig.~\ref{fig:plasmonMode}, i.e $T/T_F=0.0$ (blue solid line), $T/T_F=0.1$ (green dot-dashed line), $T/T_F=0.3$ (orange dashed line) and $T/T_F=1.0$ (red dotted line). For $q/k_F\lesssim0.74$, both the zero-temperature width from local RPA and nonlocal RPA are zero and the relative difference is undefined. This is the reason for the blue curve ending abruptly.}
\end{figure*}

Fig.~\ref{fig:dispCompare} shows a comparison between plasmon dispersions obtained from the nonlocal RPA model and the local RPA model. We calculate the dispersion for zero temperature and $T/T_F=1.0$ and it is clear from the figure that there are differences in the plasmon dispersion obtained from the two models. The local RPA (dashed lines) consistently underestimates the energy of the plasmon for a given momentum with the deviations being larger for high energy plasmons. Having obtained the plasmon dispersions, we can compute the plasmon wavelengths as a function of the energy by numerically inverting the dispersion relations\footnote{Remembering that $q=2\pi/\lambda$ we can for every $\omega$ of interest numerically find the $\lambda$ that fulfills the dispersion.}. The differences between the calculated plasmon wavelengths are shown in Fig.~\ref{fig:wavelength}. They are larger for the high energy plasmons, with a relative difference up to $20\%$. We also see that the difference between the local RPA result and the nonlocal RPA result is larger at small temperatures.\\

Fig.~\ref{fig:diffLoss} shows a comparison between the plasmon widths obtained from nonlocal RPA and local RPA. The overall trend is that for small temperatures the local RPA underestimates the width and the underestimation is larger for high energy plasmons. As the temperature increases, the local RPA becomes better and approaches the RPA result, especially for low energy plasmons.

\section{Optical scattering properties of graphene plasmon resonances\label{sec:resonancePeaks}}
In this section, we calculate the reflectance, transmittance and absorbance from the graphene-grating structure shown in Fig.~\ref{fig:setup}. The dielectric grating has a dielectric constant of $\epsilon_r=3$ and we take the dielectric rods to have the same width, $d/2$, as height (aspect ratio of 1). The dielectric rods are placed periodically along the $x$ axis with a distance of $d/2$ between them. The length of the effective unit cell of the periodicity is then $d$. In this section, we use $d\approx80$ nm, which for these parameters corresponds to $q/k_F=0.5$. Our results indicate that the aspect ratio of the dielectric rods does not play a significant role for the scattering properties. We restrict our treatment to longitudinal electric fields, meaning that the electric field is aligned as in Fig.~\ref{fig:setup}, along the direction of periodicity, and the magnetic field is thus always along the grating rods. This restriction of longitudinal fields is quite natural as we want to investigate the response to longitudinal graphene plasmons. For simplicity we also restrict our treatment to normal angles of incidence.\\

\begin{figure*}
\centering
\subfigure[]{\label{fig:refZeroK}
\includegraphics[width=0.31\textwidth]{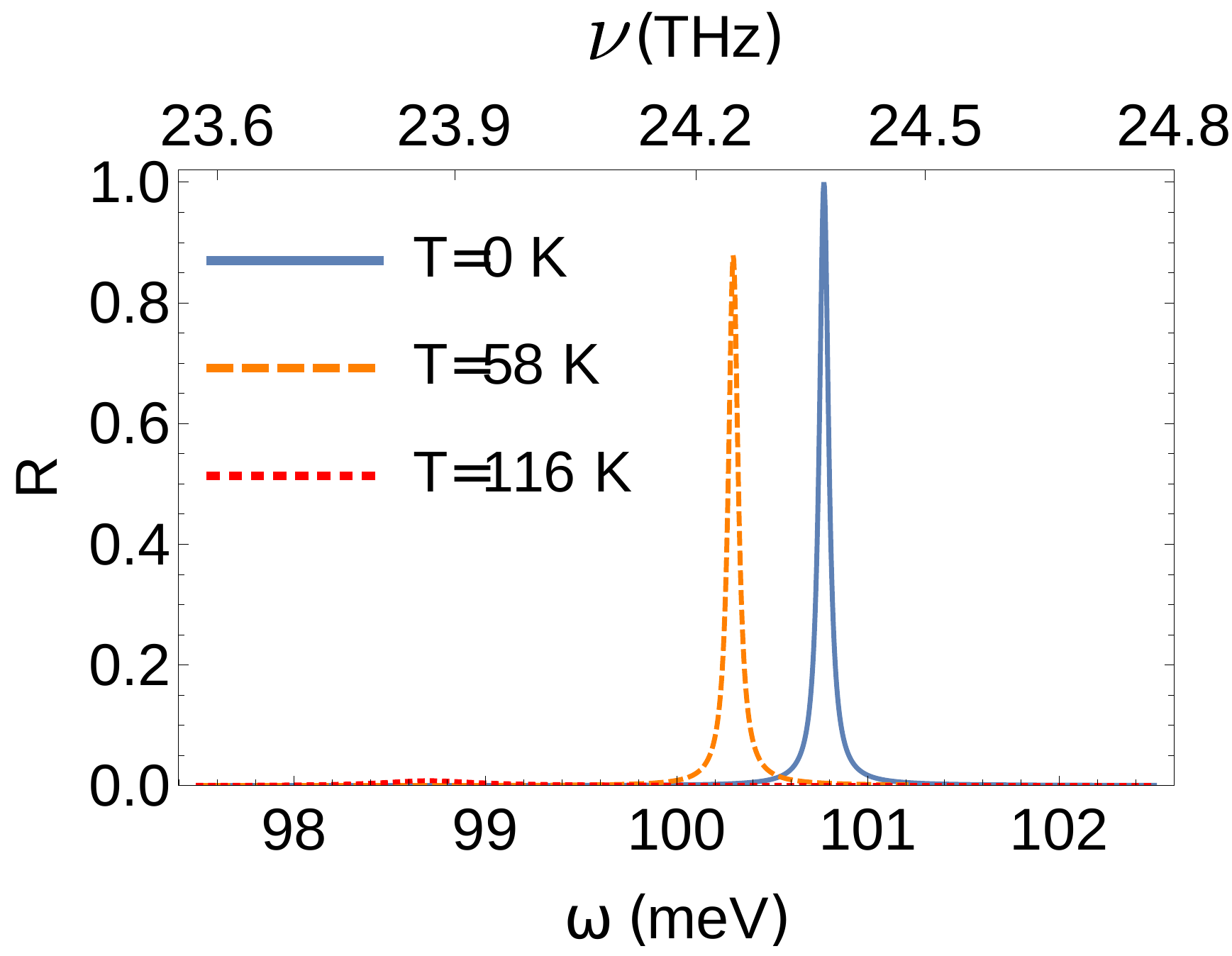}
}
\subfigure[]{\label{fig:transZeroK}
\includegraphics[width=0.31\textwidth]{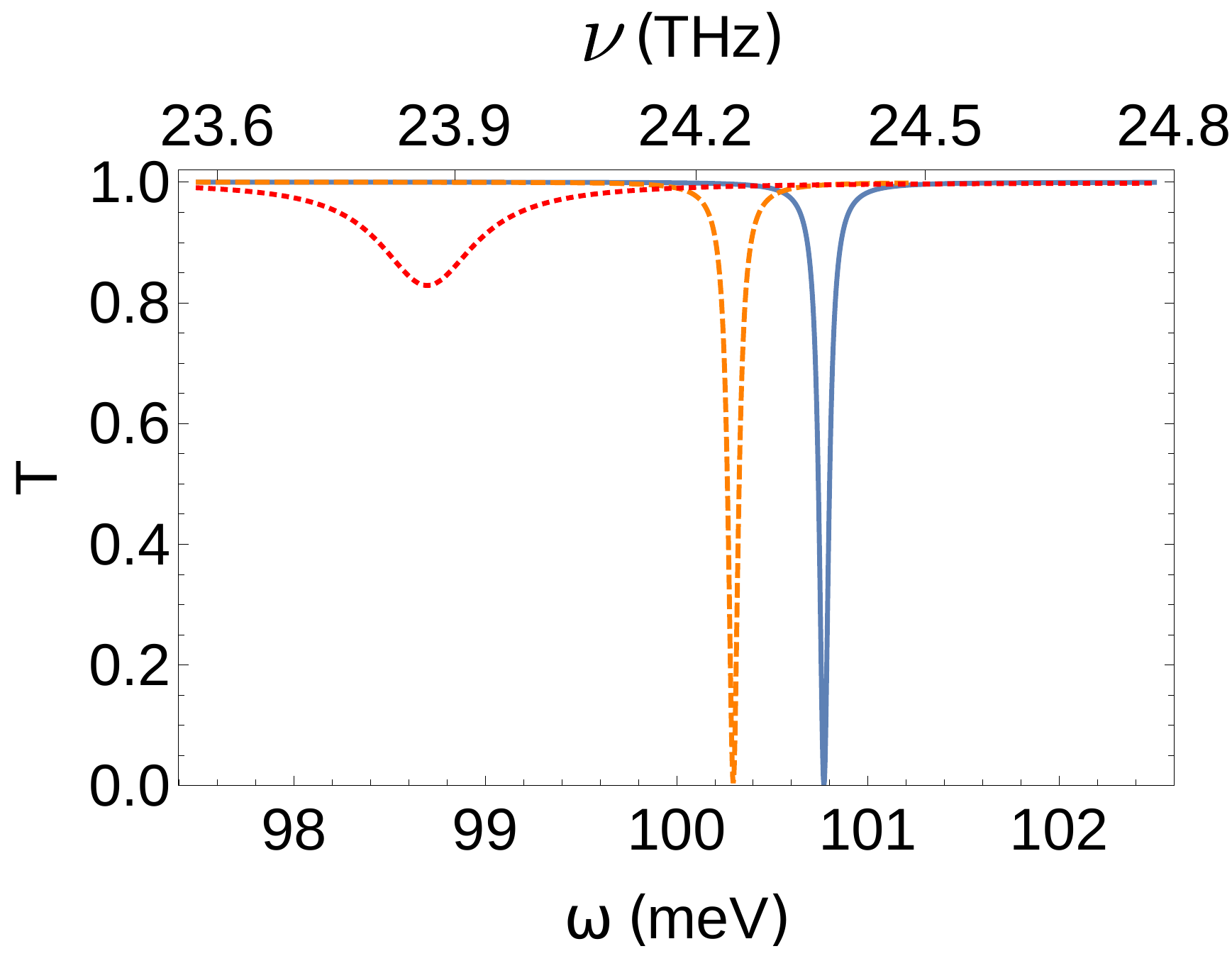}
}
\subfigure[]{\label{fig:absZeroK}
\includegraphics[width=0.31\textwidth]{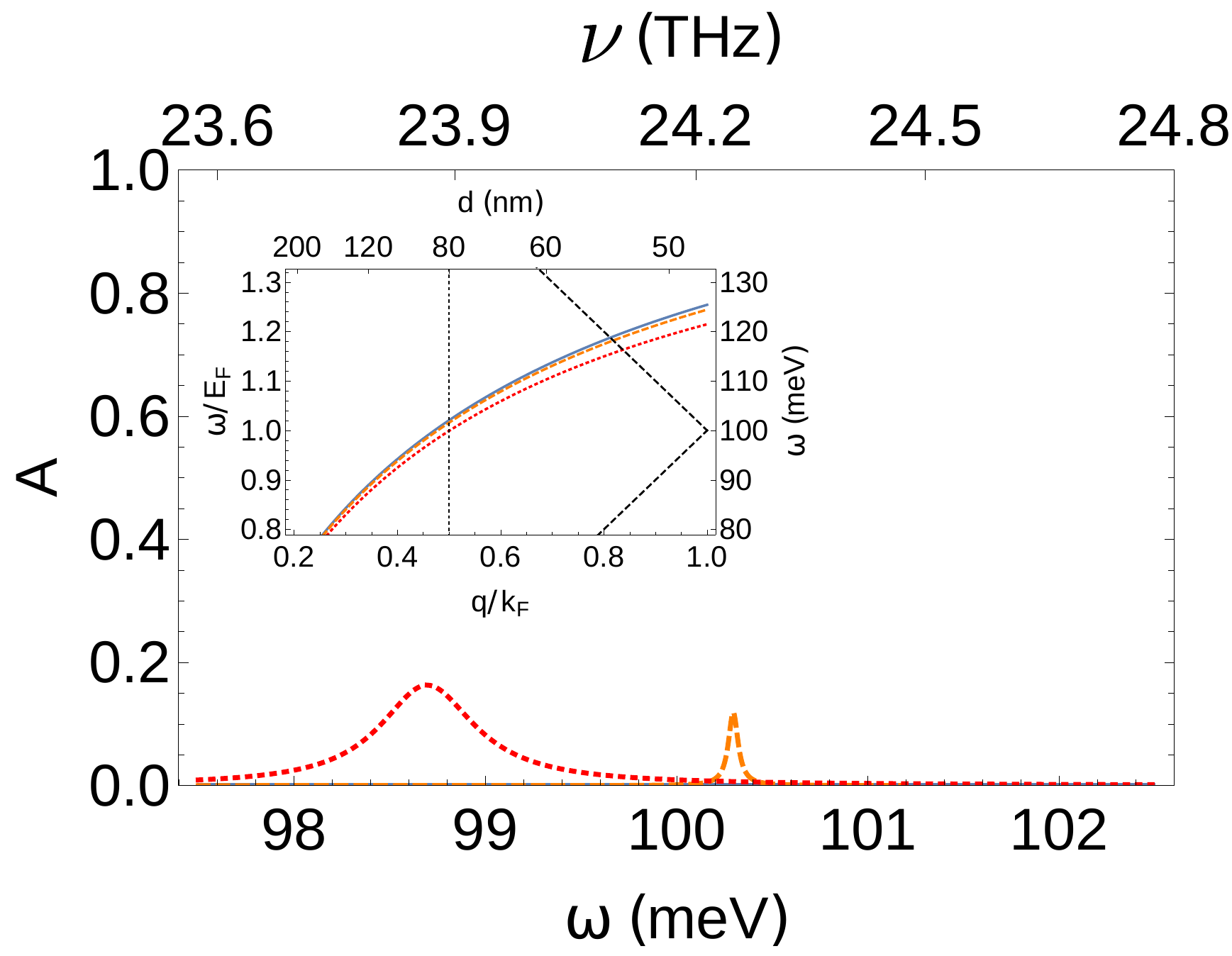}
}
\subfigure[]{\label{fig:ref}
\includegraphics[width=0.31\textwidth]{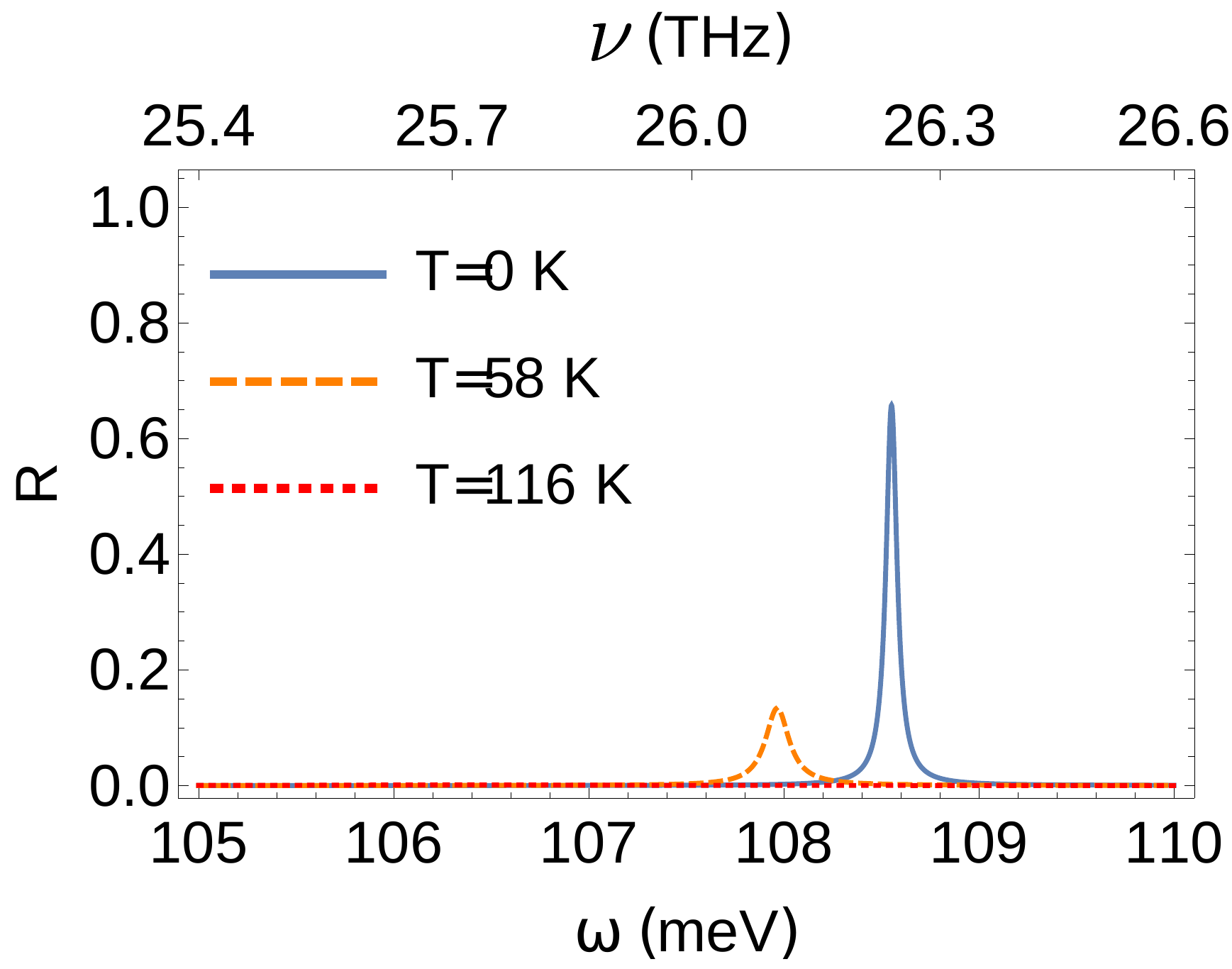}
}
\subfigure[]{\label{fig:trans}
\includegraphics[width=0.31\textwidth]{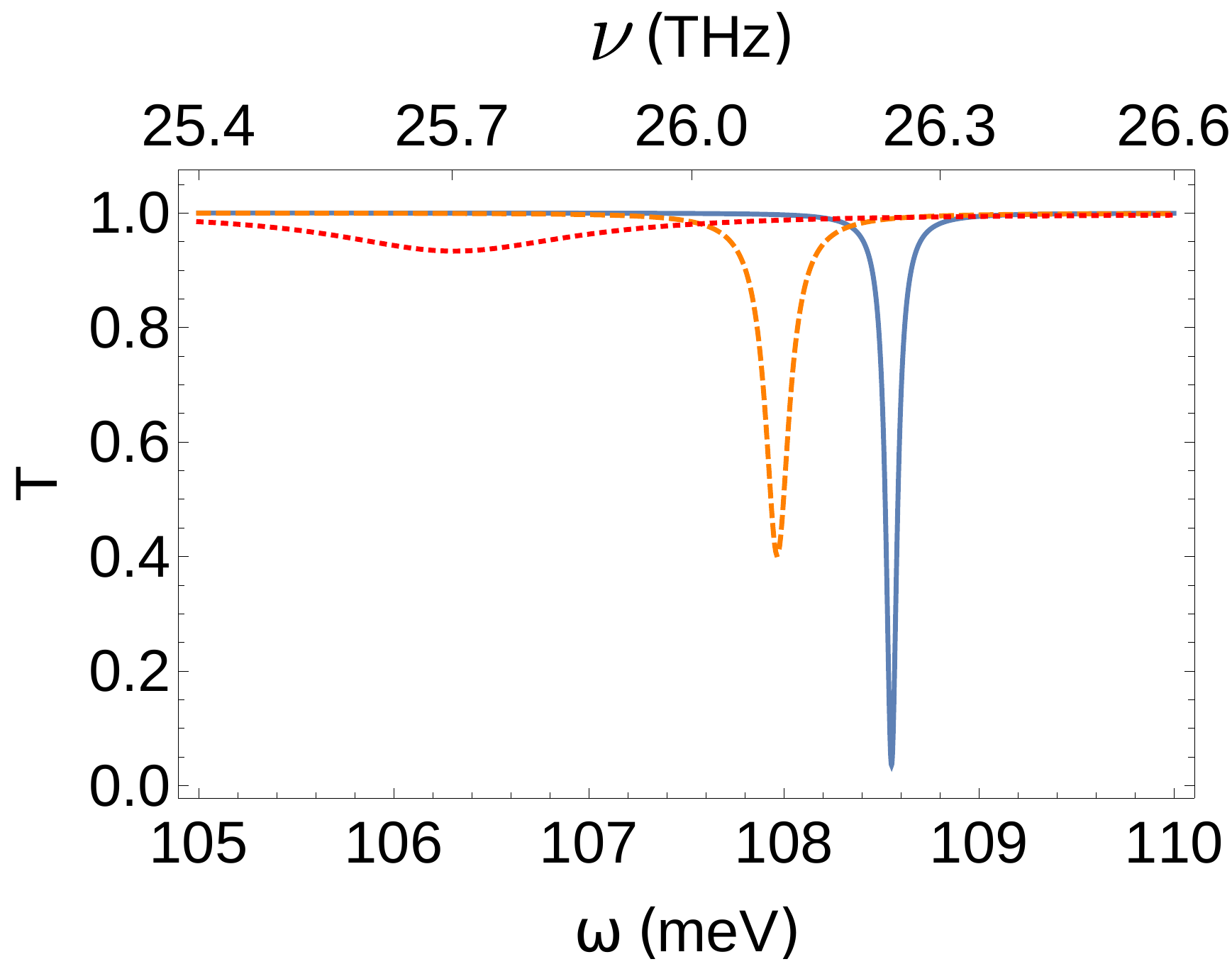}
}
\subfigure[]{\label{fig:abs}
\includegraphics[width=0.31\textwidth]{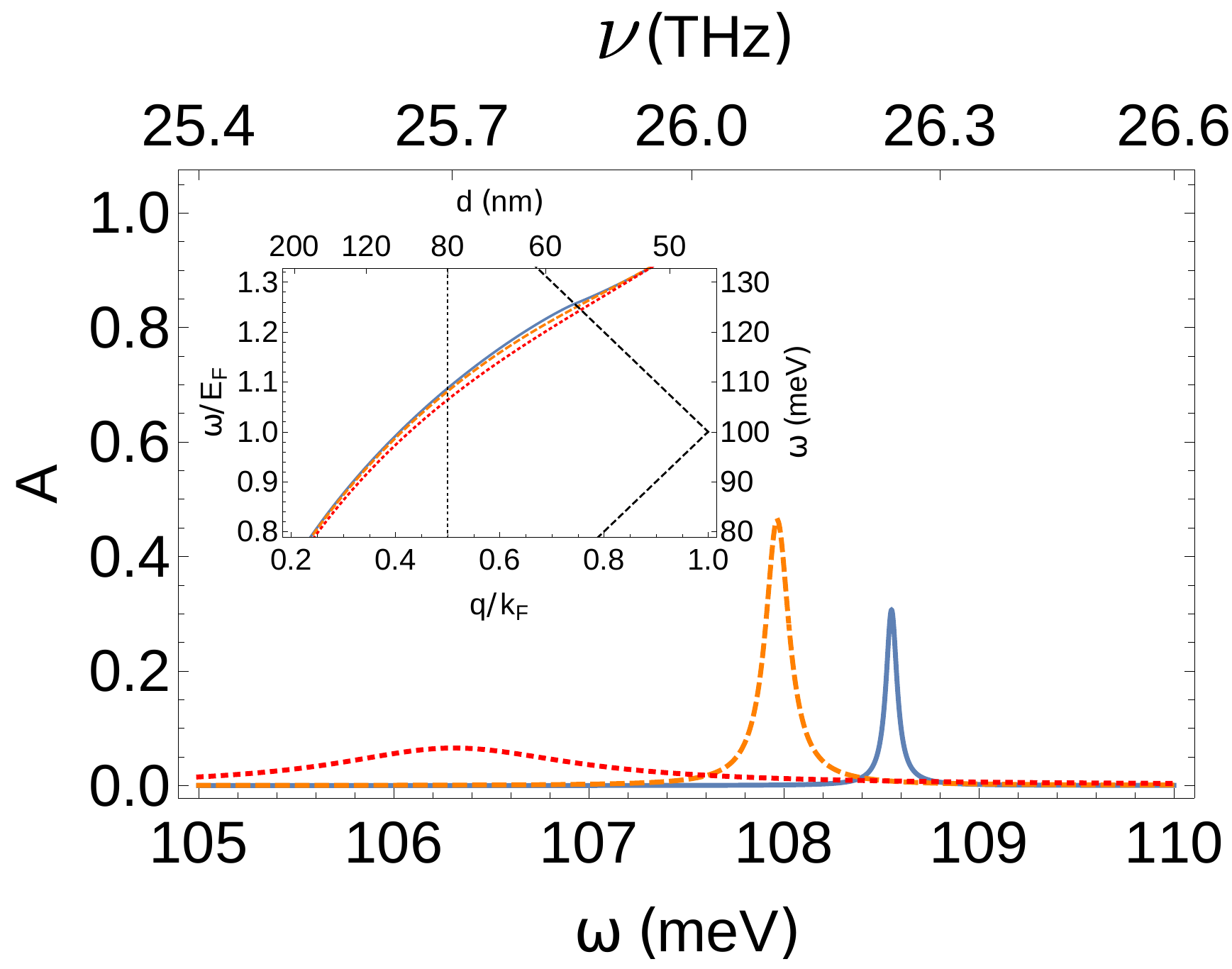}
}
\caption{\label{fig:plasmonResonance}(Color online) \textbf{Top row:} results using $\sigma(\omega)$. \textbf{Bottom row:} results using $\sigma(q,\omega)$. In all figures $T=0$ K (blue solid lines), $T=58$ K (orange dashed lines), and $T=116$ K (red dotted lines). \textbf{a), d)} Reflectance  \textbf{b), e)} Transmittance  \textbf{c), f)} Absorbance with an inset showing relevant plasmon dispersions together with a dashed line indicating the grating induced momentum $q=2\pi/d$. Note that the frequency axes are different in the top row and bottom row. 
}
\end{figure*}

We use the scattering matrix method explained in appendix \ref{app:Smatrix} and we calculate the reflectance $R$, transmittance $T$ and absorbance $A$ defined as 
\begin{align}
R&=|r|^2\\
T&=|t|^2\\
A&=1-R-T,
\end{align}
where $r$ and $t$ are the reflection and transmission amplitudes calculated in appendix \ref{app:Smatrix} and the equation for the absorbance comes from energy conservation. We set the Fermi energy of the graphene sheet to $E_F=0.1$ eV, corresponding to $n\approx0.8\cdot10^{12}/\text{cm}^2$. As was shown earlier in Fig.~\ref{fig:dispTemp}, the temperature effect on the plasmon dispersion is determined by the ratio $T/T_F$. In this section, we investigate three different temperatures, $T/T_F=0$, $T/T_F=0.05$ and $T/T_F=0.1$, which correspond to $T=0$ K, $T=58$ K and $T=116$ K, for the chosen Fermi energy. In order to compare the optical properties between $\sigma(q,\omega)$ (nonlocal RPA) and $\sigma(\omega)$ (local RPA), we do all calculations with identical parameters for both cases. The results of our calculations are shown in Fig.~\ref{fig:plasmonResonance}.\\

Fig.~\ref{fig:plasmonResonance} shows the reflectance, transmittance and absorbance of our structure as a function of frequency. The top row shows the results for the local RPA and the bottom row shows the results for the nonlocal RPA. The insets in Figs.~\ref{fig:absZeroK} and \ref{fig:abs} show the relevant plasmon dispersion together with a dashed line indicating the grating-induced momentum. In the top row of Fig.~\ref{fig:plasmonResonance} (local RPA results), we observe clear plasmon resonances and the peak positions are in good agreement with the calculated plasmon dispersions shown in the inset of Fig.~\ref{fig:absZeroK}. The plasmon response is visible in reflectance, transmittance and absorbance with the exception of the zero-temperature absorbance which is zero everywhere. This is because the local RPA conductivity at zero temperature has a vanishing real part for all energies below $2E_F$ (remember that we have no impurities or phonons in our model). Physically the vanishing real part is due to the interband transitions being Pauli blocked at zero temperature. However, for nonzero temperatures, we see that the interband transitions are allowed due to thermal smearing of the Fermi functions and this leads to non-zero absorbance. We observe that the plasmon frequency shifts towards lower frequencies when the temperature is increased. Increasing temperatures also lead to an increased broadening of the plasmon peaks together with a decrease of the reflectance, transmittance and absorbance peaks/dips. Even in the rather small $T/T_F=0.1$ results (red dotted lines), the reflectance and transmittance peaks/dips have become substantially less pronounced compared with the zero-temperature results.\\

\begin{figure*}
\centering
\subfigure[]{\label{fig:refPeakZeroK}
\includegraphics[width=0.31\textwidth]{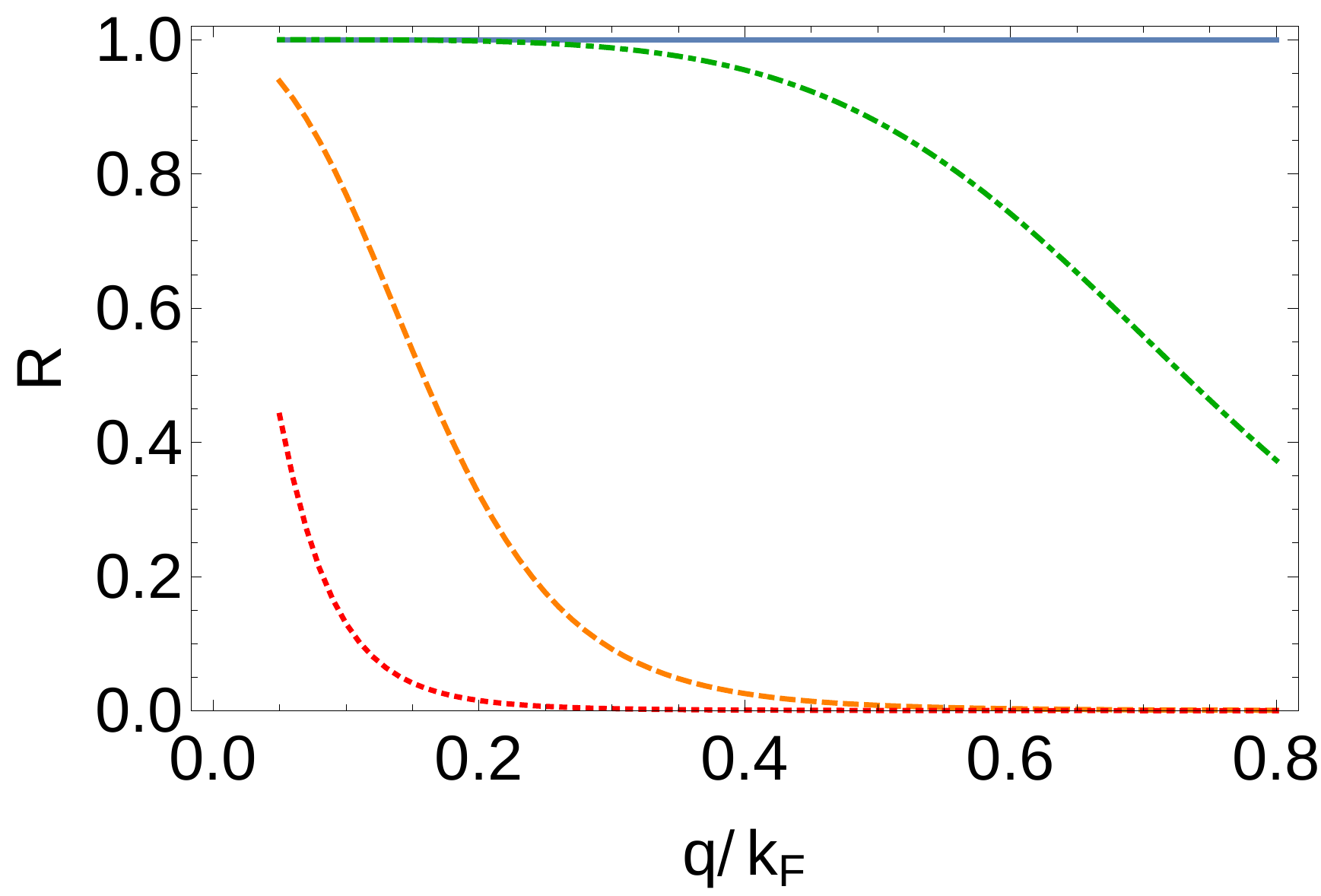}
}
\subfigure[]{\label{fig:transPeakZeroK}
\includegraphics[width=0.31\textwidth]{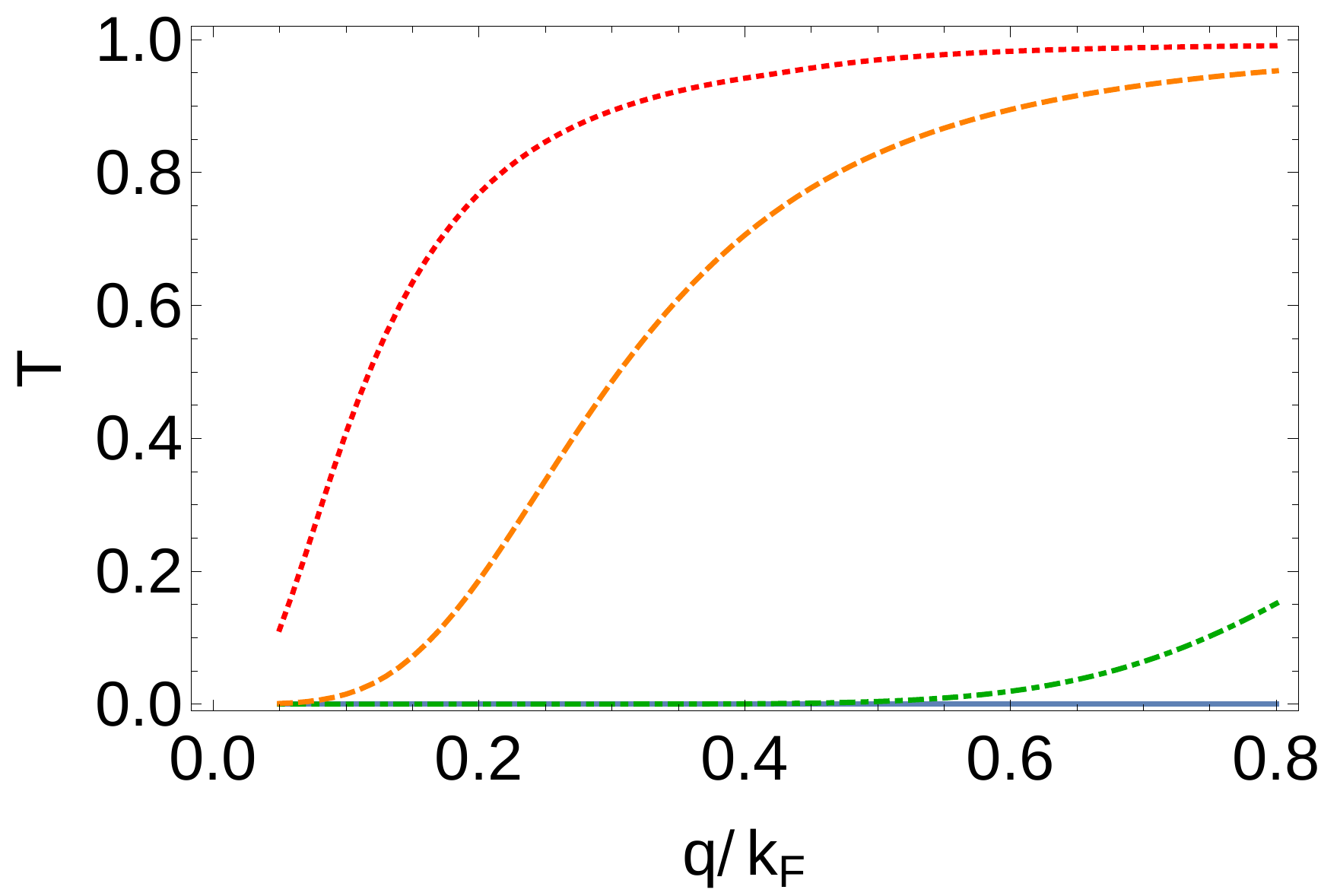}
}
\subfigure[]{\label{fig:absPeakZeroK}
\includegraphics[width=0.31\textwidth]{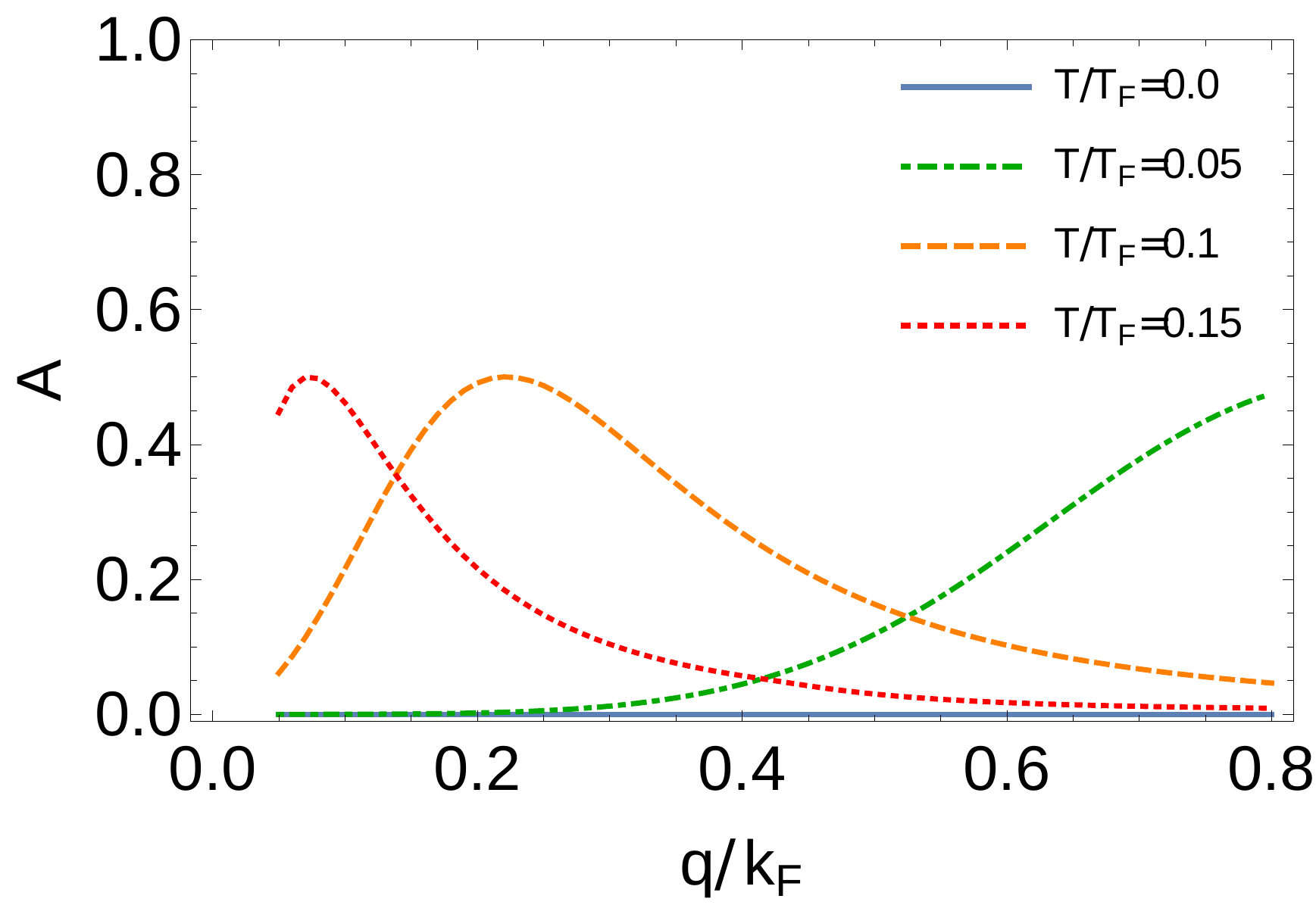}
}
\subfigure[]{\label{fig:refPeak}
\includegraphics[width=0.31\textwidth]{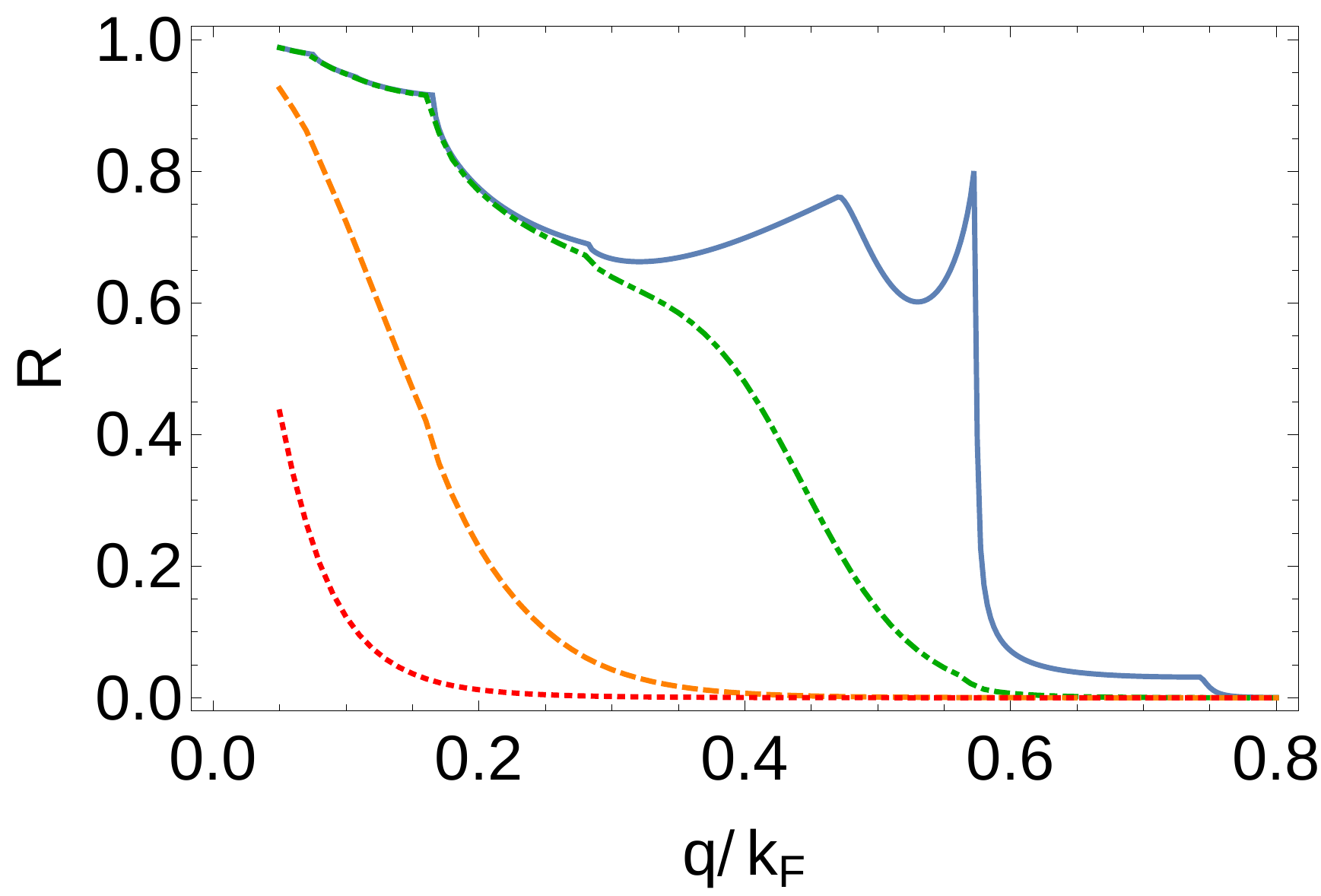}
}
\subfigure[]{\label{fig:transPeak}
\includegraphics[width=0.31\textwidth]{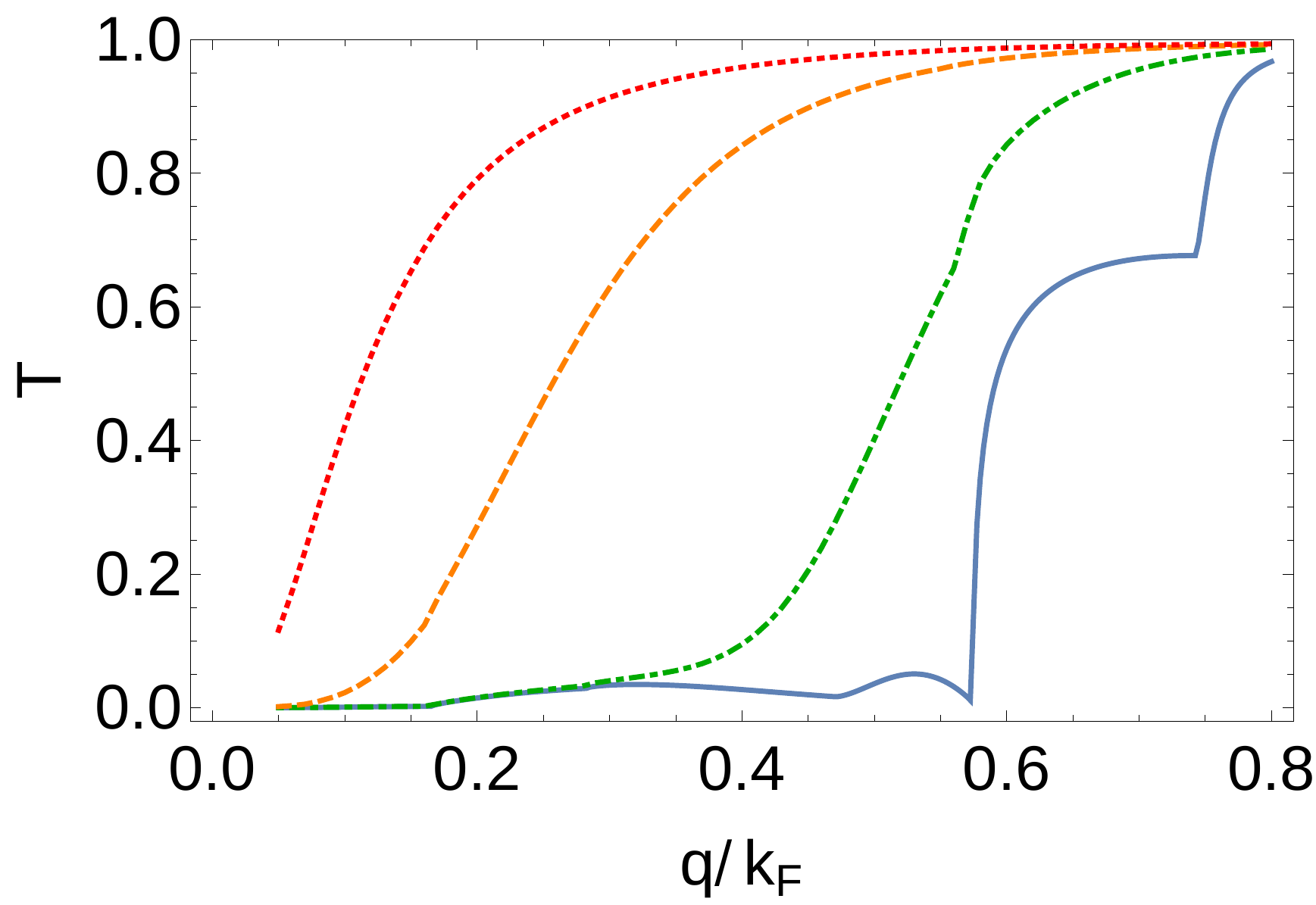}
}
\subfigure[]{\label{fig:absPeak}
\includegraphics[width=0.31\textwidth]{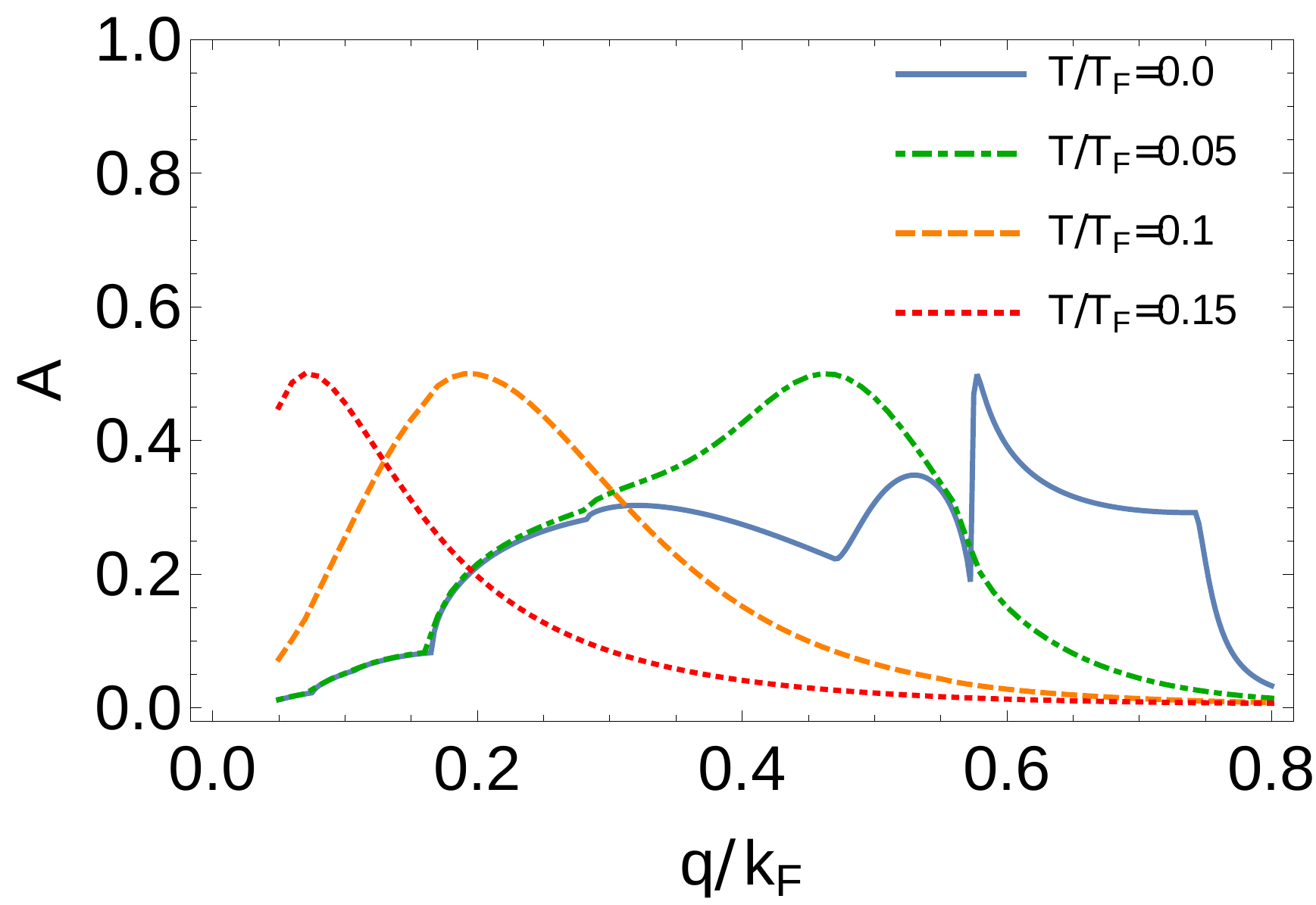}
}
\caption{\label{fig:plasmonPeak}(Color online) Scattering coefficients along the plasmon dispersion. \textbf{Top row:} results using $\sigma(\omega)$. \textbf{Bottom row:} results using $\sigma(q,\omega)$. In all figures $T/T_F=0$ (blue solid lines), $T/T_F=0.05$ (green dot-dashed lines), and $T/T_F=0.1$ (orange dashed lines) and $T/T_F=0.15$ (red dotted lines). \textbf{a,d)} Reflectance peak height \textbf{b,e)} Transmittance peak height \textbf{c,f)} Absorbance peak height.}
\end{figure*}

The bottom row of Fig.~\ref{fig:plasmonResonance} shows the calculated optical properties obtained using the nonlocal RPA conductivity. Similarly to the local RPA calculation, we observe resonance peaks in both reflectance, transmittance and absorbance. We also observe a frequency shift towards higher frequencies compared to local RPA results for the same parameters. (Note that the frequency axes are different in the top and bottom rows of Fig.~\ref{fig:plasmonResonance}.) We also see that the plasmon dispersion shifts to lower frequencies when raising the temperature. This corresponds well with the dispersions shown in the inset of Fig.~\ref{fig:abs}. Comparing the top and bottom rows, we see that using nonlocal RPA predicts smaller peaks/dips for non-zero temperatures compared to local RPA. This is clearly visible by comparing for instance the reflectance shown in the orange dashed lines in Figs.~\ref{fig:refZeroK} and \ref{fig:ref}. The difference in the reflectance for these two cases clearly illustrates the importance of taking into account the momentum dependence of the conductivity, at least in this particular case. The absorbance we obtain in the structure is rather large on resonance with the plasmon, up to $50\%$, and in contrast to the local RPA result the zero temperature absorbance is non-zero. The appearance of non-zero absorbance can be understood by considering the $q$ dependence of the nonlocal RPA conductivity. Due to the grating structure, any given frequency couples to many $q$-vectors, allowing the plasmon (also at zero temperature) to couple to the electron-hole continuum which gives rise to a non-zero real part of the conductivity. In our model this is encoded in the Fourier series expansion of the conductivity, which, for any frequency, samples many $q$ vectors, many of them outside the white triangle in Fig.~\ref{fig:dispTemp}. This effect does not appear in the local RPA, because the conductivity in that approximation is independent of $q$. The $q$ dependence in the RPA is also responsible for the overall lowering of the reflectance peaks and transmittance dips in the RPA results compared to the local RPA results. \\

We wish to once again point out that we have neglected impurities and phonons in our model and we have instead focused on the momentum dependence and the temperature effects. Of course, since the temperature effect depends on the ratio $T/T_F$, an obvious way to reduce the temperature effect is to go towards larger doping levels. We also wish to point out that the $q$ dependence introduces an additional lowering of the reflectance peaks and transmittance dips and in addition broadens the resonances compared to the local RPA. Additional broadening is an effect one would normally associate with impurities and/or electron-electron interactions and care should be taken when analyzing scattering results using the local RPA with regard to these effects.

\section{Analysis of the plasmon resonances for various grating periodicities\label{sec:gratings}}
\begin{figure}
\centering
\subfigure[]{\label{fig:totWidthZeroK}
\includegraphics[width=0.4\textwidth]{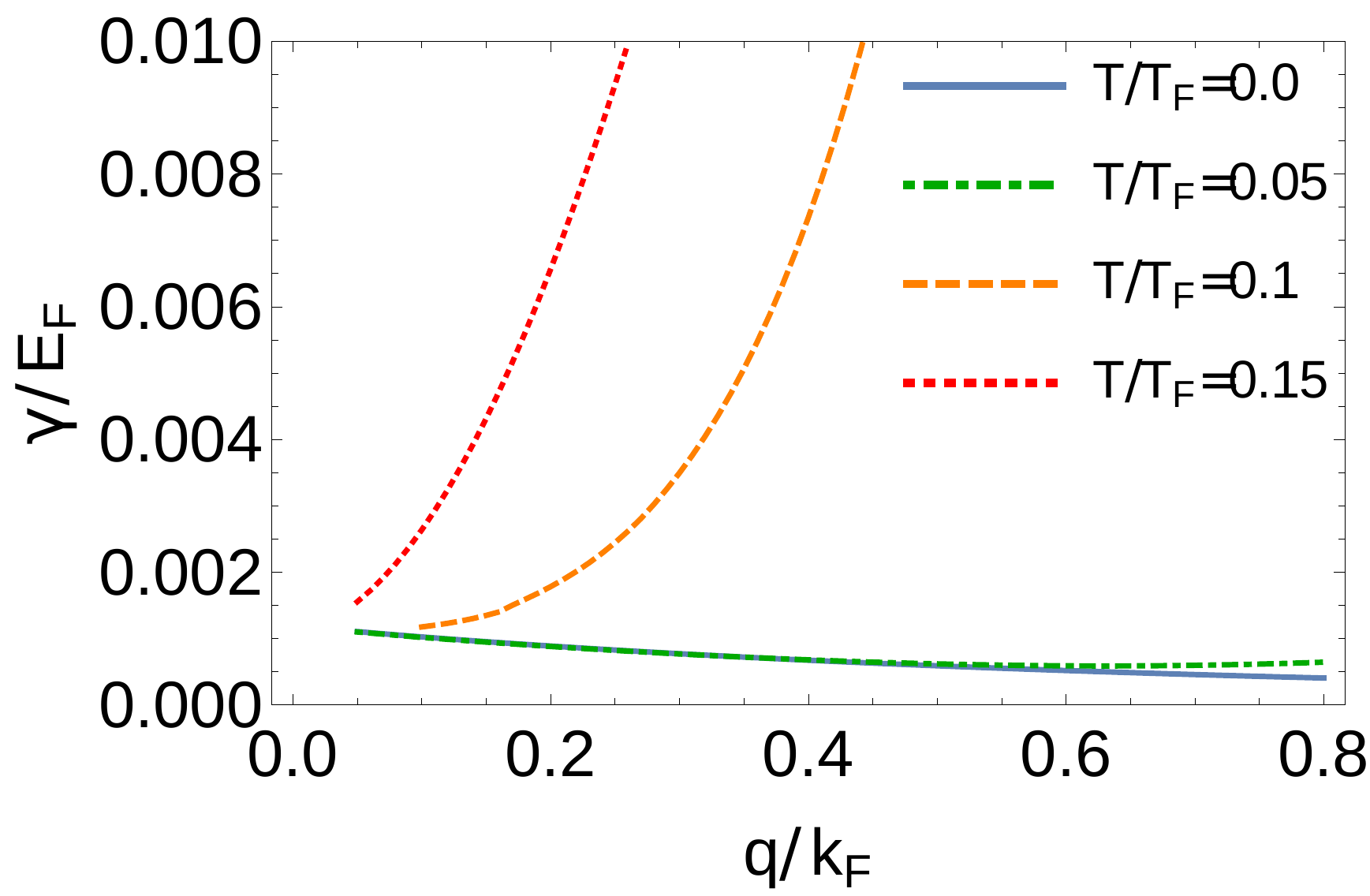}
}
\subfigure[]{\label{fig:totWidth}
\includegraphics[width=0.4\textwidth]{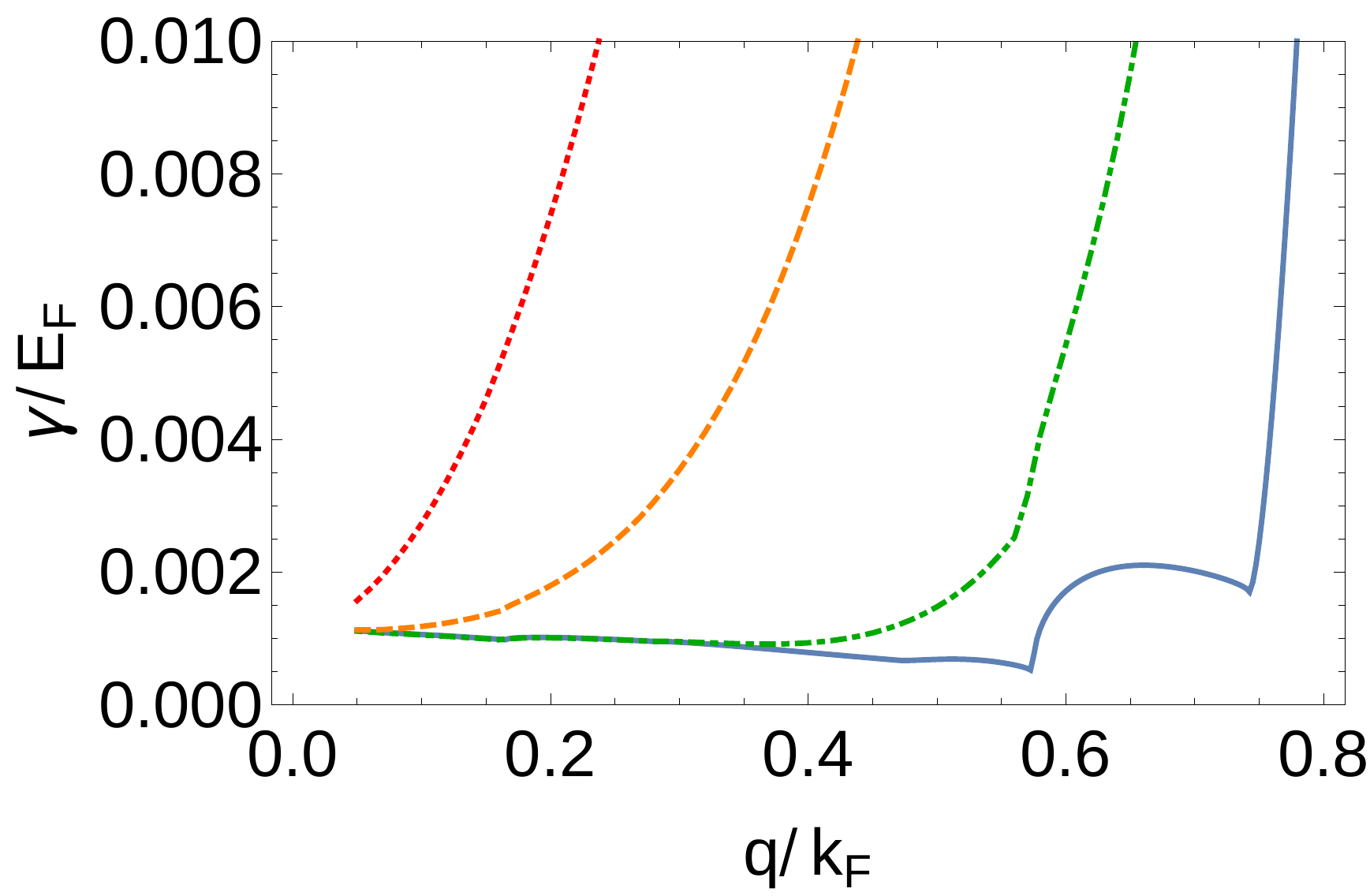}
}
\caption{(Color online) \textbf{a)} Width (FWHM) of the transmittance peaks used in Fig.~\ref{fig:transPeakZeroK}, i.e. using local RPA. \textbf{b)} Width (FWHM) of the transmittance peaks used in Fig.~\ref{fig:transPeak}, i.e using RPA. In both figures $T/T_F=0$ (blue solid lines), $T/T_F=0.05$ (green dot-dashed lines), and $T/T_F=0.1$ (orange dashed lines) and $T/T_F=0.15$ (red dotted lines).}
\end{figure}

In order to further investigate the plasmon resonance signatures and their temperature dependence, we consider several different grating periodicities. For each grating periodicity, giving rise to $q=2\pi/d$ where $d$ is the grating periodicity, we find the peak reflectance, transmittance and absorbance given by the plasmon resonance. We also calculate the width of the plasmon resonance peaks. We calculate these quantities using both the local RPA and nonlocal RPA conductivity. This allows us to quantify the behavior of the plasmon resonances as a function of grating periodicity, and by comparing the local RPA with the nonlocal RPA results, we can quantify the difference between them. We calculate these results for the temperatures $T/T_F=0.0$, $T/T_F=0.05$, $T/T_F=0.1$ and $T/T_F=0.15$. We point out that the results in section \ref{sec:resonancePeaks} were obtained using $q/k_F=0.5$. The results shown in Fig.~\ref{fig:plasmonPeak} are produced by solving the scattering problem as in section \ref{sec:resonancePeaks} for a multitude of grating periodicities and for each periodicity we find the peak reflectance, transmittance and absorbance. Also the width of the transmittance peaks was extracted and is shown in Figs.~\ref{fig:totWidthZeroK} and \ref{fig:totWidth}.\\

The top row in Fig.~\ref{fig:plasmonPeak} shows the peak value of the reflectance, transmittance and absorbance for various grating periodicities obtained with the local RPA conductivity. At zero temperature the reflectance always peaks to unity and the transmittance decreases to zero, so the absorbance must be zero. The local RPA predicts that the reflectance peak becomes smaller for larger temperatures---this is in agreement with the behavior we observed in section \ref{sec:resonancePeaks}. Also, the transmittance dip becomes smaller as the temperature increases and we see that, for any temperature, there is a particular grating period for which the absorbance is $50\%$. At a temperature of $T/T_F=0.15$ only gratings with small $q/k_F$ (large grating distance $d$) will show scattering signatures from the plasmon resonances. \\

The bottom row in Fig.~\ref{fig:plasmonPeak} shows the peak value of the reflectance, transmittance and absorbance for various grating periodicities obtained with the nonlocal RPA conductivity. We see that at zero temperature (blue solid lines) there is a non-trivial sharp behavior of the peak heights as function of $q/k_F$. This is in stark contrast to the smooth behavior at zero temperature in the local RPA results. The sharp features can be understood by again considering the $q$-dependence of the conductivity, similar to the way we understood the non-zero absorbance at zero temperature in section \ref{sec:resonancePeaks}. For a given frequency the light is coupled through the grating to many different momenta and at zero temperature the conductivity has sharp features that we are probing. 
These sharp features are quickly smeared out when going to non-zero temperatures, this can be seen in the green dot-dashed lines in Figs.~\ref{fig:refPeak} and \ref{fig:absPeak} where the sharp features only remain for small $q/k_F$. In the larger temperature results the sharp features have completely vanished and we approach the local RPA results. It is worth noting that there is a rather significant absorbance in the system due to the interaction with the single particle continuum, also at small temperatures. This is the case for both the local RPA and the nonlocal RPA results. However, the location and form of this peak as a function of grating periodicity and temperature is quite different between the two models. The absorbance has a peak value of $50\%$ and the position of this peak in terms of $q/k_F$ is temperature dependent. We note that $50\%$ saturates the theoretical upper limit for normal incidence absorbance of a thin (much thinner than the incident wavelength) structure sandwiched between two identical dielectrics established in Ref. \cite{Thongrattanasiri2012a}. In our case the graphene-grating structure is sandwiched between two half-spaces of air.\\

An important result can be obtained by comparing the local RPA results and the nonlocal RPA results. At zero temperature the results are very different but as the temperature is increased the local RPA results and the full RPA are in better agreement, at $T/T_F=0.15$ the two practically coincide. The conclusion is that nonlocal effects are only important for low temperatures compared to the Fermi temperature. We note that the Fermi temperature depends on the doping level, making finite $q$-effects more important for graphene with large doping. Conversely, the nonlocal effects are less important for small doping levels.\\

Figs.~\ref{fig:totWidthZeroK} and \ref{fig:totWidth} show the width of the transmittance peak for the various grating periodicities from Fig.~\ref{fig:plasmonPeak}. Fig.~\ref{fig:totWidthZeroK} shows the width using local RPA and Fig.~\ref{fig:totWidth} shows the width obtained using nonlocal RPA. It is clear that the widths differ between the two cases and the local RPA underestimates the resonance width for small temperatures and large $q$. For larger temperatures, the results coincide as in the results in Fig.~\ref{fig:plasmonPeak}.

\section{Conclusions}
In conclusion, we have theoretically investigated the scattering properties of monolayer graphene on a subwavelength dielectric grating and we find that such a structure, if properly tuned, can exhibit very large reflectance and transmittance signatures when the external light is on resonance with the graphene plasmon. We have compared intrinsic plasmon properties as well as scattering results obtained by using the local RPA conductivity $\sigma(q\to0,\omega)$ with results obtained by using the nonlocal RPA conductivity $\sigma(q,\omega)$. We find that small grating periodicities (large $q/k_F$) have the largest discrepancies in the optical properties obtained from the two results, indicating that in this regime the local RPA is no longer valid. A less obvious result we find is that the discrepancies between local RPA and nonlocal RPA are largest at small temperatures but as the temperature is increased the discrepancies become smaller and around $T/T_F=0.15$ the differences have almost completely vanished. For large temperatures, $T/T_F>0.15$, the local RPA conductivity is a good approximation that correctly captures the scattering properties of graphene on a subwavelength dielectric grating. \\

Furthermore, we find that the optical scattering amplitudes are heavily degraded by temperature effects, which makes grating experiments at room temperature challenging. We point out that the temperature effects depend on the ratio $T/T_F$. Thus, a possible way to reduce the degrading temperature effects is to increase the doping level and, thereby, increase the Fermi temperature. Another important aspect is that the plasmons with small $q/k_F$ (long wavelength) are less affected by the detrimental temperature effects.\\

We have also compared the intrinsic plasmon properties obtained from the local RPA and nonlocal RPA. We have shown that the calculated wavelength of the plasmons may differ up to $20\%$. For example, for a grating periodicity $d\approx80$ nm and doping level $E_F=0.1$ eV or $n\approx 0.8\cdot 10^{12}/$cm$^2$, the difference in the calculated plasmon frequency obtained from the two models is around $2$ THz. In an experiment this blue-shift may be misinterpreted if the experimental results are compared with the local RPA dispersion, but it is in fact a nonlocal effect.\\

For some applications such as solar cells, a large absorbance is beneficial and plasmonic enhancements may help improve on current technologies \cite{Atwater2010}. We point out that graphene on a subwavelength dielectric grating structure absorbs up to $50\%$ of the incident radiation close to resonance with the plasmon and that in some cases this absorbance approaches the theoretical maximum\cite{Thongrattanasiri2012a}.\\

Our results could be used to make a refractive index sensor and our narrow plasmon resonances should make it possible to detect very small refractive index changes in the dielectric environment surrounding the graphene and subwavelength dielectric grating. This refractive index sensor would be tunable since the plasmon resonance frequency is tunable by electrostatic gating.

\begin{acknowledgments}
The authors wish to thank the Knut and Alice Wallenberg foundation for financial support.
\end{acknowledgments}


\appendix

\section{Linear response theory and conductivity}\label{app:linResp}
In this paper we treat graphene within linear response theory; the unperturbed graphene Hamiltonian is \cite{katsnelson}

\begin{equation}
\hat H_0=v_F\,\vec{\sigma}\!\cdot\!\vk=\left(\begin{array}{cc}0&v_F(k_x-i k_y)\\ v_F(k_x+i k_y)&0\end{array}\right)
\end{equation}
with the graphene sheet confined to the xy-plane and $v_F$ being the Fermi velocity of graphene. We use $\hbar=1$ in the paper. We assume that inter-valley scattering is absent so both graphene valleys are independent of each other. Similarly we assume spin up and down to be independent of each other; hence spin and valley degrees of freedom are both degenerate and contribute a factor $2$ each in the final answer.\\

The real space Dyson equation is
\begin{equation}
\lbrack \varepsilon-\hat H_0\rbrack\circ\hat G_0=\hat \delta
\label{eq:dysonE0}
\end{equation}
where $\circ$ means integration/summation over all internal variables. Solving Dyson's equation for graphene we obtain the unperturbed Green's function for one valley \cite{Peres2006,Lofwander2007}
\begin{equation}
\hat G_0(\vk,\varepsilon)
=\frac{1}{2}\sum_{\lambda=\pm 1}\frac{1}{\varepsilon-\lambda v_Fk}
\left(\begin{array}{cc}
1&\lambda{\rm e}^{-i\phi_k}\\ \lambda{\rm e}^{i\phi_k}&1
\end{array}\right)
\label{G0}
\end{equation}
where $k=|\vk|$ and $\phi_k=\arg (k_x+ik_y)$. \\

We now perturb the graphene Hamiltonian with an external longitudinal electric field. For a longitudinal electric field we can write the electric field in terms of a potential as $\vec{E}=i\vec{q}\phi$. We write the perturbation Hamiltonian as (notice that this perturbation is spin and valley independent)
\begin{align}
\delta \hat{H}(x) &= -e\sum_n\phi_n(x)e^{-i\omega t+iq_nx} \pmat{1}{0}{0}{1}=\nonumber\\
&=\sum_n\frac{ie}{q_n}E_n(x)e^{-i\omega t+iq_nx}\pmat{1}{0}{0}{1}
\end{align}
where we sum over many perturbation wavelengths since we anticipate that the grating will induce a perturbation at multiple $q_n$. The Dyson equation with the perturbation becomes
\begin{equation}\label{eq:dysonth}
\lbrack \hat \varepsilon-\hat H_0-\delta \hat H \rbrack\circ \hat G=\hat \delta.
\end{equation}
In linear response theory the full Green's function is linearly perturbed by the perturbation so we write the full Green's function as $\hat G=\hat G_0 + \delta \hat G$. Inserting this ansatz in Eq.~\eqref{eq:dysonth}, to zeroth and first order in the perturbation we get
\begin{align}
(\hat \varepsilon -\hat H_0)\circ\hat G_0&=\hat \delta\\
(\hat \varepsilon -\hat H_0)\circ \delta \hat G-\delta \hat H\circ \hat G_0&=0,
\end{align}
where the higher order term in the perturbation is omitted since we are doing linear response theory. From these equations we immediately see that to linear order the perturbation to the Green's function can be written
\begin{equation}\label{eq:deltaGstart}
\delta\hat G(xt,x't') = \hat G_0\circ \delta \hat H \circ \hat G_0
\end{equation}
which means that to linear order in the perturbation the correction to the Green's function is determined by the unperturbed Green's function together with the perturbation. For fermions we can write the density response from the perturbation as \cite{fetter}
\begin{align}\label{eq:nFetter}
\langle \delta \hat n(x,t) \rangle = -i\text{Tr} \left [ \delta\hat G(xt,xt^+)  \right ] 
\end{align}
and evaluating the equal argument Green's function perturbation starting from Eq.~\eqref{eq:deltaGstart} we obtain
\begin{align}
\delta \hat G(xt,xt^+)&= \sum_n \sum_{i\epsilon_m}\int d\vk\ \hat G_0(\vk+\vec{q}_n,i\epsilon_m+i\omega_m)\times \nonumber \\
&\times \hat G_0(\vk,i\epsilon_m)\frac{ieE_n}{q_n}e^{iq_nx-i\omega t}.
\end{align}
where $\vec{q}_n=q_n\hat x$. Inserting this expression into Eq.~\eqref{eq:nFetter}, performing the Matsubara summation over $i\epsilon_m$ \cite{mahan} and doing the trace in sublattice space we obtain
\begin{align}
\langle \delta\hat n(x,t)&=\sum_n \langle \delta \hat n_n(x,t) \rangle\\
\langle \delta \hat n_n(x,t) \rangle &=i g_sg_v \sum_{\lambda\lambda'}\int d\vk\ \frac{f_{\lambda,k}-f_{\lambda',k+q_n}}{i\omega_m-\lambda' v_F|k+q_n|+\lambda v_F k}\times\nonumber\\
&\times\frac{1}{2} (1+\lambda \lambda' \cos (\phi_{k+q_n}-\phi_k))\frac{eE_n}{q_n}e^{iq_nx-i\omega t}
\end{align}
where $f_{k,\lambda}$ is the Fermi function and we have inserted $g_s$ and $g_v$ for spin and valley degeneracy. The continuity equation
\begin{align}
e \partial_t \langle n_n(x,t) \rangle &= -\nabla \cdot \langle j_n(x,t) \rangle \Rightarrow\\
\Rightarrow \frac{e\omega}{q_n}\langle n_n(q_n,\omega)\rangle &=\langle j_n(q_n,\omega) \rangle
\end{align}
gives us the current as
\begin{align}
\langle j(q,\omega) \rangle&=\sum_n \langle j_n(q,\omega) \rangle \\
\langle j_n(q,\omega) \rangle &=\sigma(q,\omega)\delta(q-q_n)E_n\\
\sigma(q,\omega)&=\frac{ig_sg_v\omega e^2}{q^2}\times\nonumber\\
 &\times\sum_{\lambda\lambda'}\int d\vk\ \frac{f_{\lambda,k}-f_{\lambda',k+q}}{i\omega_m-\lambda' v_F|\vk+\vec{q}|+\lambda v_F k}\times\nonumber\\
&\times\frac{1}{2} (1+\lambda \lambda' \cos (\phi_{k+q}-\phi_{k}))
\end{align}
These equations tell us that for each mode $q_n$ of the perturbation there is an associated current component at the same wave number and the total current is obtained by adding all these contribution. Performing analytic continuation on the conductivity $(i\omega_m \rightarrow \omega +i\eta)$ and defining the polarizability we obtain the final answer

\begin{align}
\sigma(q,\omega)&=\frac{i e^2\omega}{q^2}\Pi(q,\omega)\label{eq:sigmaPi}\\
\Pi(q,\omega)&=\lim_{\eta\to 0^+}\frac{g_sg_v}{2}\int d\vec{k}\sum_{\lambda,\lambda'}\frac{f_{\lambda,k}-f_{\lambda',k'}}{\omega +i\eta -\lambda'v_F |\vk+\vec{q}| + \lambda v_F k}\times\nonumber\\
&\times(1+\lambda\lambda'\cos\phi_{k,k'}).\label{eq:PiFinal}
\end{align}
where $\vec{k}'=\vk+\vec{q}$, $\Pi(q,\omega)$ is the polarizability (density-density correlation function) and $\phi_{k,k'}=\phi_{k+q}-\phi_k$.\\ 

The only thing left to do is to compute the polarizability-integral in Eq.~\ref{eq:PiFinal}. This was computed in Refs \cite{Hwang2007,Wunsch2006} for zero temperature and this was later generalized to finite temperatures in Ref. \cite{Ramezanali2009}. In the paper we refer to the result in equations \ref{eq:sigmaPi} and \ref{eq:PiFinal} as the nonlocal RPA result or simply the RPA result.\\

In this paper we also use the conductivity at zero wave vector where the answer simplifies to \cite{Falkovsky2007,Falkovsky2008}
\begin{align}\label{eq:locCondOne}
\sigma(\omega)&=\Re[ \sigma(\omega)] + i \Im [\sigma(\omega)]\\
\Re [\sigma(\omega)]&=\frac{e^2}{4}G(\omega/2)\\
\Im[\sigma(\omega)]&=\frac{e^2}{4}\bigg ( \frac{8T}{\omega \pi}\ln \left [ 2\cosh \left ( \frac{\mu}{2T}\right ) \right ]+\nonumber\\
 &+\frac{4\omega}{\pi}\int_0^{\infty} \frac{G(x)-G(\omega/2)}{\omega^2-4x^2} dx \bigg )\\
G(x)&=\frac{\sinh (x/T)}{\cosh (\mu/T)+\cosh (x/T)}\label{eq:locCondEnd}
\end{align}
where $\mu$ is the chemical potential and $T$ is the temperature. In the paper we refer to equations \eqref{eq:locCondOne}-\eqref{eq:locCondEnd} as the local RPA result or the local RPA conductivity.

\section{Scattering matrix method}\label{app:Smatrix}
We use a scattering matrix method \cite{Li2003} to investigate the optical properties of the subwavelength dielectric grating-graphene structure. This is a convenient way to solve electromagnetic scattering problems because it allows the problem to be subdivided into smaller pieces, if necessary, and then recombined to obtain the solution to the full problem. In this appendix we use units where $\mu_0=\varepsilon_0=c=\hbar=1$ and $e^2=4\pi\alpha$ where $\alpha\approx1/137$ is the fine structure constant.\\

\begin{figure}[h!]
\centering
\includegraphics[width=0.4\textwidth]{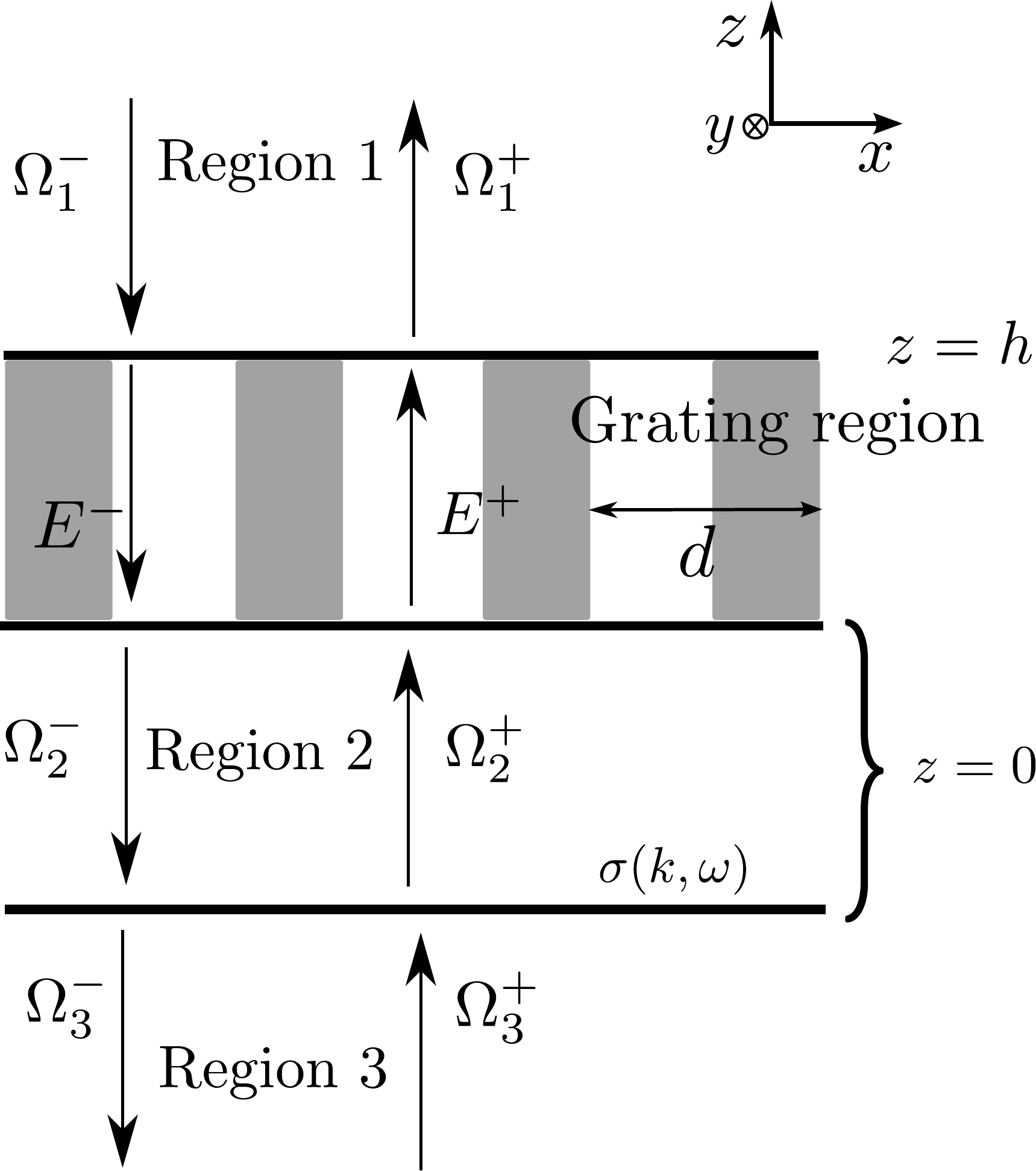}
\caption{\label{fig:gratingSetup}Figure showing the different regions considered in the scattering problem. Region 2 is the thin air film and the shaded areas represent the periodic dielectric function.}
\end{figure}

Our structure consists of several distinct regions, see Fig.~\ref{fig:gratingSetup}. The upper region, region 1, is a semi-infinite half space from which the EM radiation is incident, underneath is the periodic sub-wavelength dielectric grating (subwavelength dielectric grating). Region 2 is an infinitely thin air film and region 3 is a semi-infinite half space that fills the bottom half space into which the transmitted radiation is propagating. The graphene is here considered as an infinitely thin conducting sheet underneath the subwavelength dielectric grating and appears as a modified boundary condition when matching the waves between the thin air film and the semi-infinite half space.\\

The scattering matrix method is based on Fourier decomposing the electric and magnetic fields and by matching the boundary conditions we relate the Fourier amplitudes between the different regions. The periodicity is set by the grating region, here we take the dielectric function to be a periodic step function between the values $\epsilon_r=1$ and $\epsilon_r=3$ in the $x$ direction, with total periodicity set by the distance $d$. Putting the dielectric step of width $d/2$ in the middle the Fourier series expansion of the dielectric function is
\begin{align}
\varepsilon(x)&=\frac{\varepsilon_0+\varepsilon_1}{2}-\sum_{j\in\text{odd}}a_je^{ik_jx}\\
a_j&=\frac{\varepsilon_1-\varepsilon_0}{\pi j}\ \sin \left (\frac{\pi j}{2} \right )\\
k_j&=\frac{2\pi j}{d}.\label{eq:kGrating}
\end{align}
We restrict our treatment to normal incidence with the incident electric field in the $x$ direction and the incident magnetic field in the $y$ direction. In our setup this makes the incident electric field parallel to the periodic axis and we call it longitudinal to the periodicity. It is important to point out that $E_x$ and $H_y$ are enough to describe propagating modes. However, since we will consider also evanescent modes we must also include an $E_z$ component in our treatment.

\subsection*{Fields in free space}
In order to find expressions for free space fields we need two of Maxwell's equations, namely
\begin{align}
\nabla \times \vec{E} + \partial_t \vec{B}&=0\label{eq:nablaCrossE}\\
\nabla \times \vec{H}-\partial_t \vec{D}&=0\label{eq:nablaCrossH},
\end{align}
where the space and time dependence of the fields is not written explicitly. By applying a rotation to one of the above equations and then inserting the other in the resulting equation we can obtain the dispersion relation for a Maxwell field as

\begin{equation}\label{eq:MaxwellDisp}
\omega^2=\beta^2+q^2
\end{equation}
where $q$ is now in the $x,y$ direction which span the graphene plane, and $\beta$ is in the $z$-direction perpendicular to the graphene plane. Notice that a mode is evanescent in the $z$-direction if $\omega^2<q^2$ since this forces $\beta$ to be imaginary.\\

Using equations  \eqref{eq:nablaCrossE} and \eqref{eq:nablaCrossH} with harmonic time dependence we obtain (remember that $\mu_0=\epsilon_0=1$ so that $\vec{E}=\vec{D}$ and $\vec{H}=\vec{B}$)
\begin{align}
\nabla\times\vec{E}&=i\omega\vec{H}\label{eq:HfromE}\\
\nabla \times \vec{H}&= -i\omega\vec{E}\label{eq:EfromH}.
\end{align}
Writing these equations out in their components (for $E_x$, $E_z$ and $H_y$) we get 
\begin{align}
\partial_z E_x -\partial_x E_z &= i\omega H_y\\
\partial_x H_y &=-i\omega E_z\\
\partial_z H_y &=i\omega E_x
\end{align}
and using the second equation to eliminate $E_z$ in the first one we get 
\begin{equation}\label{eq:ExHy}
i\omega H_y + \frac{i}{\omega}\partial_x^2 H_y = \partial_z E_x.
\end{equation}

We now write out the electric and magnetic fields as Fourier sums

\begin{align}
E^{\pm}_x=\sum_j \Omega^{\pm}_{j}\ e^{ik_jx}\ e^{\pm i\beta_j z}\\
H^{\pm}_y=\sum_j \Lambda^{\pm}_{j} \ e^{ik_jx}\ e^{\pm i\beta_j z}
\end{align}
where $+$ $(-)$ denotes waves propagating in positive (negative) $z$ direction and from Eq.~\eqref{eq:MaxwellDisp}
\begin{equation}\label{eq:betaj}
\beta_j^2=\omega^2-k_j^2.
\end{equation} Now we plug the above Fourier expansions into Eq.~\eqref{eq:ExHy} and multiply by $e^{-ik_jx}$ and integrate over $x$ to project out equations for the amplitudes obtaining
\begin{equation}
\left ( \omega -  \frac{k_j^2}{\omega}  \right ) \Lambda^{\pm}_j = \pm  \beta_j \Omega^{\pm}_j.
\end{equation}
Now, inverting the prefactor on the left and remembering that $\beta_j$ is defined by Eq.~\eqref{eq:betaj} we obtain a relationship between the magnetic field amplitudes $\Lambda_j$ and electric field amplitudes $\Omega_j$ as
\begin{equation}
\Lambda_j^{\pm}=\pm\frac{1}{\sqrt{1-k_j^2/\omega^2}}\Omega_j^{\pm}.
\end{equation}
This relationship means that we can fully determine the magnetic field if we know the electric field and vice-versa. We choose to work with the electric field amplitudes $\Omega_j$'s as our fundamental objects. We write
\begin{equation}
\Lambda_j^{\pm}=\pm T_{0,j}\Omega_j^{\pm}.
\end{equation}
where 
\begin{equation}
T_{0,j}=\frac{1}{\sqrt{1-k_j^2/\omega^2}}.
\end{equation}
 \\

We now introduce $\vec{E}_m=(....E_{-n}....E_0....E_n....)^T$ which is a vector containing all electric field amplitudes and similarly $\vec{H}_m=(....H_{-n}....H_0....H_n....)^T$ for the magnetic field. In the same manner we write the unknown Fourier expansion coefficients $\vec{\Omega}^{\pm}_m=(....\Omega^{\pm}_n....\Omega^{\pm}_0....\Omega_n^{\pm}....)^T$ and we may then write the matrix relationship

\begin{equation}\label{eq:freeSpace}
\pvec{\vec{E}_m}{\vec{H}_m}=\pmat{S_0}{S_0}{T_0}{-T_0}\pvec{\vec{\Omega}^+_m}{\vec{\Omega}^-_m}
\end{equation}
where $T_0$ is a diagonal matrix with $T_{0,j}$ on the diagonal and $S_0$ is the unit matrix. Reintroducing the spatial dependence in Eq.~\eqref{eq:freeSpace} we get the expression for the fields in free space as 
\begin{equation}\label{eq:fieldDecomp}
\pvec{\vec{E}_m(x,z)}{\vec{H}_m(x,z)}=\pmat{S_0}{S_0}{T_0}{-T_0}\pvec{\vec{\Omega}_m^+(x,z)}{\vec{\Omega}_m^-(x,z)}
\end{equation}
where 
\begin{align}
&\vec{\Omega}^{\pm}_m(x,z)=\nonumber\\
&=(...\Omega^{\pm}_{-m}\ e^{i(k_{-m}x \pm \beta_{-m} z)}..\Omega^{\pm}_0e^{\pm i\omega z }..\Omega^{\pm}_{m}\ e^{i(k_mx \pm \beta_m z)}...)^T.
\end{align}
We point out that the electric field in equations \eqref{eq:freeSpace} and \eqref{eq:fieldDecomp} is the component along the $x$ axis and the magnetic field is the field along the $y$ axis.\\

\subsection*{Wave matching at the graphene interface}
In order to compute the S matrix for the graphene sheet we must wave match across a conducting interface with the boundary conditions \cite{jackson}

\begin{align}\label{eq:BC1}
\vec{H}_y(0^-)-\vec{H}_y(0^+)&=j_x=\sigma \vec{E}_x(0^{\pm})\\
\label{eq:BC2}
\vec{E}_x(0^-)-\vec{E}_x(0^+)&=0
\end{align}
where $+$ denotes the fields above the graphene sheet and $-$ the fields below. In terms of field amplitudes, see Fig.~\ref{fig:gratingSetup}, these boundary conditions become

\begin{align}
\Omega_2^-+\Omega_2^+ &= \Omega_3^-\\
\Omega_2^- - \Omega_2^+ &=\left (S_0+ T_0^{-1}\sigma \right ) \Omega_3^-.
\end{align}
By adding and subtracting these two equations we get 

\begin{align}
2\Omega_2^-&= \left ( 2S_0+T_0^{-1}\sigma  \right ) \Omega_3^+\\
2\Omega_2^+&= -T_0^{-1}\sigma \Omega_3^-
\end{align}
from which we can solve
\begin{align}
\Omega_3^-&= \left ( S_0+\frac{1}{2}T_0^{-1}\sigma  \right )^{-1} \Omega_2^-\\
\Omega_2^+&= -\left ( 2S_0+T_0^{-1}\sigma  \right )^{-1}T_0^{-1}\sigma\Omega_2^-.
\end{align}
Also, since the graphene scattering problem has inversion symmetry around $z=0$ the reflection and transmission amplitudes from the other side are identical. This gives us the final S-matrix for a graphene sheet as

\begin{align}\label{eq:S2final}
S_{\text{graphene}}&=\pmat{2M}{-M\ T_0^{-1}\sigma}{-M\ T_0^{-1}\sigma}{2M}\\
M&=\left ( 2S_0+T_0^{-1}\sigma  \right )^{-1}.
\end{align}
where $\sigma$ is a diagonal matrix with $\sigma(k_j,\omega)$ on the diagonal.\\

\subsection*{Fields in the grating}
Following Ref. \cite{Li2003} we rewrite Maxwell's equations inside the grating into an eigenproblem for $E_x$.\\

We use equations \eqref{eq:nablaCrossH} and \eqref{eq:nablaCrossE} (no currents inside the grating) with harmonic time dependence giving us
\begin{align}
\nabla \times \vec{E} = i\omega\vec{H}\\
\nabla \times \vec{H}= -i\underbrace{\varepsilon(x)\vec{E}}_{\vec{D}}
\end{align}
and the difference between this and the free space expression is the periodic dielectric function $\varepsilon(x)$. We only have the components $E_x$, $E_z$ and $H_y$, and our system is translationally invariant in the $y$-direction so all fields are $y$ independent. Inserting this in the above equations gives us the following three equations
\begin{align}
\partial_zE_x-\partial_x E_z=i\omega H_y\\
\partial_x H_y = -i\omega \varepsilon(x)E_z\\
-\partial_z H_y = -i\omega \varepsilon(x)E_x.
\end{align}
Using the second of these we get an expression for $E_z$ as 
\begin{equation}\label{eq:EzfromHy}
E_z=\frac{i\varepsilon^{-1}(x)}{\omega}\partial_x H_y
\end{equation}
and plugging this back into the other two we obtain

\begin{align}
\partial_z E_x &=i\omega H_y +\frac{i}{\omega}\partial_x \left ( \varepsilon^{-1}(x)\partial_x H_y \right )\\
\partial_z H_y &=i\omega \varepsilon(x)E_x.
\end{align}

We now Fourier expand all fields, including $\varepsilon^{-1}(x)$, giving us the following expressions
\begin{align}
\sum_j \partial_z E_je^{ik_jx}&=i\omega\sum_jH_je^{ik_jx}\nonumber\\
&-\frac{i}{\omega}\sum_{j,l}\varepsilon^{-1}_jE_l(k_j+k_l)k_le^{i(k_j+k_l)x}\\
\sum_j \partial_z H_je^{ik_jx}&=i\omega\sum_{j,l}\varepsilon_jE_le^{i(k_j+k_l)x}
\end{align}
where the the fields are expanded with $k_j$ given by Eq.~\eqref{eq:kGrating}.\\

In order to project out equations for the Fourier coefficients we apply $\int dx\ e^{-ik_nx}$ to both equations giving us
\begin{align}
\partial_zE_n&=i\omega\sum_l\delta_{n,l}H_l-\frac{i}{\omega}\sum_l\varepsilon^{-1}_{n-l}k_nk_lH_l\\
\partial_zH_n&=i\omega\sum_l \varepsilon_{n-l}E_l
\end{align}
which can be recast into matrix form as
\begin{align}
\partial_z\vec{E}_n=T_1^{n,l}\vec{H}_l\label{eq:EtoH}\\
\partial_z\vec{H}_n=T_2^{n,l}\vec{E_l}
\end{align}
where as before $\vec{E}_n=(....E_{-m}....E_0....E_m....)^T$ and $\vec{H}_n=(....H_{-m}....H_0....H_m....)^T$ are column vectors containing the field amplitudes and
\begin{align}
T_1^{l,n}&=i\omega\delta_{n,l}-\frac{i}{\omega}\varepsilon^{-1}_{n-l}k_nk_l\\
T_2^{l,n}&=i\omega\varepsilon_{n-l}
\end{align}
are matrices. Combining these equations we obtain the eigenproblem for $\vec{E}_l$ as
\begin{equation}\label{eq:eigenProb}
\partial_z^2\vec{E}_l=\underbrace{T_1^{l,n}T_2^{n,m}}_{P^{l,m}}\vec{E}_m.
\end{equation}
Now, to solve this we introduce 
\begin{equation}\label{eq:EtildeDef}
\widetilde{E}_m=S^{-1}_a\vec{E}_l\ \Leftrightarrow \ \vec{E}_l=S_a\widetilde{E}_m
\end{equation}
where $S_a$ is a matrix which has the eigenvectors of $P$ as its columns. $\widetilde{E}_m$ is also a vector, but we have dropped the arrow for brevity. Multiplying Eq.~\eqref{eq:eigenProb} with $S_a^{-1}$ from the left and inserting a unity on the RHS we get
\begin{equation}
\partial_z^2 S_a^{-1}\vec{E}_l=S_a^{-1}P\underbrace{S_aS^{-1}_a}_{S_0}\vec{E}_m
\end{equation}
where $S_0$ denotes the unit matrix. According to Eq.~\eqref{eq:EtildeDef} this can be written in terms of $\widetilde{E}_m$ as 
\begin{equation}
\partial_z^2 \widetilde{E}_l=\underbrace{S_a^{-1}PS_a}_{D}\widetilde{E}_m
\end{equation}
and the matrix D contains the eigenvalues of P on its diagonal. \footnote{It is true in general that $A^{-1}XA=Y$ where Y contains the eigenvalues of $X$ on the diagonal if $A$ is constructed as the matrix that has the eigenvectors of $X$ as its columns.} For convenience we define a new diagonal matrix $$\gamma^2=-D$$ where the diagonal matrix $\gamma$ has $\gamma_m$ on its diagonal. We can now write the solution for $\widetilde{E}(z)$ as 
\begin{equation}
\widetilde{E}_m(z)=\widetilde{E}_m^+e^{i\gamma_mz}+\widetilde{E}_m^-e^{-i\gamma_mz}
\end{equation}
and according to Eq.~\eqref{eq:EtildeDef} the proper electric field is
\begin{equation}
\vec{E}_m(z)=S_a \left ( \widetilde{E}_m^+e^{i\gamma_m z}+\widetilde{E}_m^-e^{-i\gamma_m z} \right )
\end{equation}
and according to Eq.~\eqref{eq:EtoH} the magnetic field is
\begin{equation}
\vec{H}_m(z)=\underbrace{iT_1^{-1}S_a\gamma_m}_{T_a}\left ( \widetilde{E}_m^+e^{i\gamma_m z}-\widetilde{E}_m^-e^{-i\gamma_m z}  \right )
\end{equation}
from which we define $$T_a=iT_1^{-1}S_a\gamma.$$

We now define new $z$ dependent vectors
\begin{align}
\widetilde{E}^+_m(z)&=(....\widetilde{E}^+_{-m}e^{i\gamma_{-m}z}....\widetilde{E}^+_0e^{i\gamma_{0}z}....\widetilde{E}^+_me^{i\gamma_{m}z}....)^T\nonumber\\
\widetilde{E}^-_m(z)&=(....\widetilde{E}^-_{-m}e^{-i\gamma_{-m}z}....\widetilde{E}^-_0e^{-i\gamma_{0}z}....\widetilde{E}^-_me^{-i\gamma_{m}z}....)^T\nonumber
\end{align}
and we can finally write down the matrix expression
\begin{equation}
\pvec{\vec{E}_m(z)}{\vec{H}_m(z)}=\pmat{S_a}{S_a}{T_a}{-T_a}\pvec{\widetilde{E}^+_m(z)}{\widetilde{E}^-_m(z)}
\end{equation}
which is the solution for the fields inside the grating.\\

\subsection*{Wave matching in the grating}
To find the S matrix of the grating we start with the matching across the plane $z=h$, see Fig.~\ref{fig:gratingSetup}, which gives
\begin{equation}\label{eq:S1eq1}
\pmat{S_0}{S_0}{T_0}{-T_0}\pvec{\vec{\Omega}^+_1}{\vec{\Omega}^-_1}=\pmat{S_a}{S_a}{T_a}{-T_a}\pvec{\widetilde{E}^+(h)}{\widetilde{E}^-(h)}
\end{equation}
and we also know how to propagate the field inside the grating from $z=h$ to $z=0$, namely
\begin{equation}\label{eq:S1eq2}
\pvec{\widetilde{E}^+(h)}{\widetilde{E}^-(h)}=\pmat{e^{i\gamma h}}{0}{0}{e^{-i\gamma h}}\pvec{\widetilde{E}^+(0)}{\widetilde{E}^-(0)}.
\end{equation}
The matching across $z=0$ into the thin air film is done similarly
\begin{equation}\label{eq:S1eq3}
\pmat{S_0}{S_0}{T_0}{-T_0}\pvec{\vec{\Omega}^+_2}{\vec{\Omega}^-_2}=\pmat{S_a}{S_a}{T_a}{-T_a}\pvec{\widetilde{E}^+(0)}{\widetilde{E}^-(0)}.
\end{equation}
Now we insert Eq.~\eqref{eq:S1eq2} into \eqref{eq:S1eq1} and in the resulting equation we plug the expression for $(\widetilde{E}^+(0),\widetilde{E}^-(0))^T$ obtained from Eq.~\eqref{eq:S1eq3}. Then we make sure to remove the numerically unstable behavior from the resulting equation by multiplying so that we only have $e^{i\gamma h}$ (and not $e^{-i\gamma h}$). The result is
\begin{align}
\pmat{S_a^{-1}+T_a^{-1}T_0}{S_a^{-1}-T_a^{-1}T_0}{e^{i\gamma h}(S_a^{-1}-T_a^{-1}T_0)}{e^{i\gamma h}(S_a^{-1}+T_a^{-1}T_0)}&\pvec{\vec{\Omega}^+_1}{\vec{\Omega}^-_1}=\nonumber \\
=\pmat{e^{i\gamma h}(S_a^{-1}+T_a^{-1}T_0)}{e^{i\gamma h}(S_a^{-1}-T_a^{-1}T_0)}{S_a^{-1}-T_a^{-1}T_0}{S_a^{-1}+T_a^{-1}T_0}&\pvec{\vec{\Omega}^+_2}{\vec{\Omega}^-_2}
\end{align}
and defining new variables we write this as
\begin{align}
\pmat{p}{m}{e^{i\gamma h}m}{e^{i\gamma h}p}\pvec{\vec{\Omega}^+_1}{\vec{\Omega}^-_1}= \pmat{e^{i\gamma h}p}{e^{i\gamma h}m}{m}{p}\pvec{\vec{\Omega}^+_2}{\vec{\Omega}^-_2}.
\end{align}
where 
\begin{align}
p=S_a^{-1}+T_a^{-1}T_0\\
m=S_a^{-1}-T_a^{-1}T_0.
\end{align}

We can now rewrite this on S-matrix form 
\begin{align}
\pvec{\Omega_2^-}{\Omega_1^+}=S_{\text{grating}}\pvec{\Omega_1^-}{\Omega_2^+}
\end{align}
where
\begin{align}\label{eq:S1final}
S&_{\text{grating}}=\\
&=\pmat{Qe^{i\gamma h}(p-m p^{-1}m)}{Q(e^{i\gamma h}mp^{-1}e^{i\gamma h}p-m)}{Q(e^{i\gamma h}mp^{-1}e^{i\gamma h}p-m)}{Qe^{i\gamma h}(p-m p^{-1}m)}\nonumber
\end{align}
with
\begin{align}
Q&=(p-e^{i\gamma h}mp^{-1}e^{i\gamma h}m)^{-1}.
\end{align}
~\\

\subsection*{Combining two S-matrices}

Finally we now seek the total S-matrix of both the grating structure and the graphene interface, i.e the S-matrix obtained by combining equations \eqref{eq:S2final} and \eqref{eq:S1final}. Writing 
\begin{align}
S_{\text{grating}}&=\pmat{t_1}{r_1}{r_1}{t_1}\\
S_{\text{graphene}}&=\pmat{t_2}{r_2}{r_2}{t_2}
\end{align}
and remembering how they are defined, see Fig.~\ref{fig:gratingSetup}, we can write down the relationships
\begin{align}
\pvec{\Omega_{2}^-}{\Omega_1^+}&=S_{\text{grating}}\ \ \pvec{\Omega_{1}^-}{\Omega_{2}^+}\\
\pvec{\Omega_{3}^-}{\Omega_{2}^+}&=S_{\text{graphene}}\pvec{\Omega_{2}^-}{\Omega_{3}^+}
\end{align}
and the S-matrix we seek is the total S-matrix that relates
\begin{align}
\pvec{\Omega_{3}^-}{\Omega_{1}^+}=S\pvec{\Omega_{1}^-}{\Omega_{3}^+}
\end{align}
which means we must eliminate $\Omega_2^{\pm}$. Eliminating both \footnote{Since all entities in the equation are matrices the order plays a role in the algebraic expression obtained. The expressions are of course equivalent, however this might not be obvious at first glance. The answer we give represent the one the authors believe to be most compact.} we obtain the total S-matrix as
\begin{align}\label{eq:Stot}
S_{\text{tot}}=\pmat{t_2q_1t_1}{r_2+t_2r_1q_2t_2}{r_1+t_1r_2q_1t_1}{t_1q_2t_2}
\end{align}
with
\begin{align}
q_1=(S_0-r_1r_2)^{-1}\\
q_2=(S_0-r_2r_1)^{-1}.\label{eq:StotFinal}
\end{align}
Equations \ref{eq:Stot}-\ref{eq:StotFinal} represent the final answer in our scattering matrix method. The answer is a block matrix where each block is $N\times N$, $N$ being the number of Fourier modes used in the expansion of the fields and the grating dielectric function. Since the S-matrix relates incoming fields with outgoing fields it is now simple to multiply the S-matrix with the incoming field amplitudes and read out the reflected and transmitted fields. We point out that in this paper we work in a regime such that there is only one propagating mode in the reflected field and one propagating mode in the transmitted field. All other modes are evanescent and as such they do not carry any energy away from the structure. To investigate the power flow in the direction perpendicular to the graphene, the $z$ direction, it is thus only necessary to determine one field component amplitude for the reflected field and one for the transmitted field. This is the reflection amplitude $r$ and the transmission amplitude $t$ used in the paper to calculate the reflectance, transmittance and absorbance.

%

\end{document}